\documentclass{emulateapj}
\usepackage{amssymb,amsmath,graphicx,longtable,subfigure,wrapfig}
\usepackage{rotating}
\usepackage{tablefootnote}
\usepackage[colorlinks,linkcolor=blue,anchorcolor=green,citecolor=blue]{hyperref}

\begin{document}

\def\func#1{\mathop{\rm #1}\nolimits}
\def\unit#1{\mathord{\thinspace\rm #1}}

\title{A Monte Carlo approach to magnetar-powered transients: II.
Broad-lined type Ic supernovae not associated with GRBs}
\author{L. J. Wang\altaffilmark{1,2}, Z. Cano\altaffilmark{3,4}, S. Q. Wang%
\altaffilmark{5,6,7}, W. K. Zheng\altaffilmark{7}, L. D. Liu%
\altaffilmark{5,6}, J. S. Deng\altaffilmark{2}, H. Yu\altaffilmark{5,6}, Z.
G. Dai\altaffilmark{5,6}, Y. H. Han\altaffilmark{2}, D. Xu\altaffilmark{2},
Y. L. Qiu\altaffilmark{2}, J. Y. Wei\altaffilmark{2}, B. Li\altaffilmark{5,1}%
, and L. M. Song\altaffilmark{1}}

\begin{abstract}
Broad-lined type Ic supernovae (SNe Ic-BL) are a subclass of rare core
collapse SNe whose energy source is debated in the literature. Recently a
series of investigations on SNe Ic-BL with the magnetar (plus $^{56}$Ni)
model were carried out. Evidence for magnetar formation was found for the
well-observed SNe Ic-BL 1998bw and 2002ap. In this paper we systematically
study a large sample of SNe Ic-BL not associated with gamma-ray bursts. We
use photospheric velocity data determined in a homogeneous way. We find that
the magnetar+$^{56}$Ni model provides a good description of the light curves
and velocity evolution of our sample of SNe Ic-BL, although some SNe (not
all) can also be described by the pure-magnetar model or by the
two-component pure-$^{56}$Ni model (3 out of 12 are unlikely explained by
two-component model). In the magnetar+$^{56}$Ni model, the amount of $^{56}$%
Ni required to explain their luminosity is significantly reduced, and the
derived initial explosion energy is, in general, in accordance with neutrino
heating. Some correlations between different physical parameters are
evaluated and their implications regarding magnetic field amplification and
the total energy reservoir are discussed.
\end{abstract}


\affil{\altaffilmark{1}Astroparticle Physics,
Institute of High Energy Physics,
Chinese Academy of Sciences, Beijing 100049, China; wanglingjun@ihep.ac.cn}

\affil{\altaffilmark{2}Key Laboratory of Space Astronomy and Technology,
National Astronomical Observatories,
Chinese Academy of Sciences, Beijing 100012, China; wjy@nao.cas.cn}

\affil{\altaffilmark{3}Instituto de Astrof\'isica de Andaluc\'ia (IAA-CSIC),
Glorieta de la Astronom\'ia s/n, E-18008, Granada, Spain.}

\affil{\altaffilmark{4}Juan de la Cierva Fellow.}

\affil{\altaffilmark{5}School of Astronomy and Space Science, Nanjing
University, Nanjing 210093, China; dzg@nju.edu.cn}

\affil{\altaffilmark{6}Key Laboratory of
Modern Astronomy and Astrophysics (Nanjing University),
Ministry of Education, Nanjing 210093, China}

\affil{\altaffilmark{7}Department of Astronomy, University of California,
Berkeley, CA 94720-3411, USA}

\section{Introduction}

Over the past two decades, the discovery of broad-lined type Ic supernovae %
\citep[SNe Ic-BL; see][for the classification of known SNe]{Filippenko97}
and superluminous SNe (SLSNe) has greatly enlarged the family of known
core-collapse SNe (CCSNe). The association between long-duration gamma-ray
burst (GRB) 980425 and its spectroscopically associated Ic-BL SN~1998bw %
\citep{Galama98,Patat01}, i.e. the so-called GRB-SN connection %
\citep[e.g.,][]{Woosley06,Cano17}, ignited interest in these energetic and
rare type of stripped-envelope CCSNe.

To date, the luminosity of most, if not all, GRB-SNe and SNe Ic-BL could be
explained by radioactive heating arising from energy deposition from the
radioactive decay of nickel and cobalt, which is nucleosynthesized during
the explosion, into their daughter products \citep{Cano16}. However, it
appears that the luminosity of many SLSNe cannot be adequately explained in
this scenario, and alternative energy sources have been proposed. As a
consequence, it is now usually assumed that at least a subclass of SLSNe,
type Ic SLSNe, are powered by millisecond magnetars 
\citep{Kasen10,Woosley10,Chatzopoulos12,Inserra13,Nicholl14,Metzger15,Mosta15,
WangWang15,Dai16,Kashiyama16} although there is evidence for interaction
between ejecta and circumstellar medium \citep{Yan15, WangLiu16, Chen17} at
late times.

For SNe Ic-BL, shortcomings of one-dimensional (1D) $^{56}$Ni model %
\citep[e.g.,][]{Iwamoto00,Nakamura01a} stimulated the suggestion for a
two-component $^{56}$Ni model \citep{Maeda03}. In this model it is assumed
that the ejecta are composed of two components, the outer fast-moving
component (jet) and the inner slow-moving component (core). The former is
responsible for the bright peak of the light curve, while the latter is
responsible for the late-time exponential decay. This model is very useful
for providing a better description of the ejecta structure and has been very
successful in reproducing the luminosity of most SNe Ic-BL.

Recently, the application of the magnetar model to SNe Ic-BL was considered %
\citep{Cano16,WangHan16,WangWang16,WangYu17}, which are built upon the
pioneering works of \cite{Ostriker71}, \cite{Wheeler00}, and \cite%
{Thompson04}. The proposition of the improved magnetar model %
\citep{WangWang16}, which takes into account the photospheric recession and
acceleration of the ejecta by the spinning-down magnetar, provides an
opportunity to examine the magnetar model against SNe Ic-BL in a
self-consistent way. It was shown that the spin-down of the magnetar will
lose a small fraction of its rotational energy to its light curve %
\citep{WangHan16}, while the remaining fraction is transferred into the
kinetic energy of the ejecta. Evidence for the formation of stable magnetars
following the explosions of SNe Ic-BL was subsequently found by \cite%
{WangYu17}. Such a model can also naturally account for the mysterious
origin of the huge kinetic energies of SNe Ic-BL \citep{WangHan16}.

The discovery of relativistic SNe Ic-BL, 2009bb and 2012ap, through their
bright late-time radio emission %
\citep{Bietenholz10,SoderbergChakraborti10,Chakraborti11,Chakraborti15}
places the magnetar model on a more solid ground because such events require
central engines to accelerate a tiny fraction of the ejecta to
quasi-relativistic velocities \citep{Margutti14}. Actually there is a
continuous distribution of various types of CCSNe on the kinetic energy
profile of the ejecta \citep{Soderberg06}. The relativistic SNe Ic-BL lie in
between ordinary SNe Ibc and energetic GRBs and are similar to the
sub-energetic GRBs, e.g. GRB~100316D \citep{Margutti13} and GRB 140606B %
\citep{Cano15}. This may indicate that similar engines were operating in
sub-energetic GRBs and SNe 2009bb and 2012ap.

Based on the above findings, here we test the hypothesis that all of SNe
Ic-BL are powered by magnetars. Under such hypothesis, we assessed the
validity of the derived fitting parameters and consider the statistical
characteristics of SNe Ic-BL. Despite the paucity of observed SNe Ic-BL, the
accumulation of such events has reached a level where a meaningful
statistical results can start to be obtained. It is therefore very timely to
confront a larger sample ($N=11$) of SNe Ic-BL with the magnetar model.

To determine the uncertainties in the fitting parameters, \cite{WangYu17}
developed a Markov chain Monte Carlo\ (MCMC) code on the basis of the
magnetar model. This code was applied to SLSNe Ic \citep{Liu17} to minimize
the total errors arising from fitting the model to the SN light curves, and
evolution of photospheric velocity and temperature, if available. In this
paper we focus on the SNe Ic-BL not associated with GRBs. In what follows we
use the words \textquotedblleft SNe Ic-BL" to indicate SNe Ic-BL not
associated with GRBs except when specifically mentioned otherwise.

The structure of this paper is as follows. In Section \ref{sec:data} we
present the data available in the literature, along with a detailed analysis
on the uncertainties of the data. Then in Section \ref{sec:result} we
present our fitting results of the known SNe Ic-BL. Section \ref{sec:dis}
discusses the implications of the results. Particularly, Section \ref%
{sec:T_Neb} discusses the estimation of the appearance of nebular features
by early light curve modeling; Section \ref{sec:cor} discusses the
correlations between the derived parameters; Section \ref{sec:alter}
discusses the possibility of alternative models to interpret the light
curves and velocity evolution of some SNe. A summary is given in Section \ref%
{sec:conclusion}.

\section{SN sample and data analysis}

\label{sec:data}

\begin{table*}[tbph]
\caption{The SNe Ic-BL sample}
\label{tbl:sample}
\begin{center}
\begin{tabular}{cccccccc}
\hline\hline
SN & \multicolumn{2}{c}{References} & Extinction\tablenotemark{a} & Colour%
\tablenotemark{b} & \multicolumn{3}{c}{Luminosity distance} \\ 
\cline{2-3}\cline{6-8}
\multicolumn{1}{l}{} & Light curve & Velocity & corrected & used & in
referenced paper\tablenotemark{c} $\left( \unit{Mpc}\right) $ & method%
\tablenotemark{d} & adopted \tablenotemark{e}$\left( \unit{Mpc}\right) $ \\ 
\hline
\multicolumn{1}{l}{1997ef} & I00 & M16 & None & $-$ & $52.3$ & C & $50.6$ \\ 
\multicolumn{1}{l}{2002ap} & T06 & M16 & GH & $B-I$ & $7.94$ & L & $9.22$ \\ 
\multicolumn{1}{l}{2003jd} & V08 & M16 & GH & $B-R$ & $78$ & C & $84.3$ \\ 
\multicolumn{1}{l}{2007bg} & Y10 & M16 & G & $B-R$ & $147$ & C & $157.0$ \\ 
\multicolumn{1}{l}{2007ru} & S09 & M16 & G & $V-R$ & $67.6$ & C & $69.2$ \\ 
\multicolumn{1}{l}{2009bb} & P11 & M16 & GH & $V-R$ & $40$ & L & $40.68$ \\ 
\multicolumn{1}{l}{2010ah/PTF10bzf} & M13,C11 & C11 & G & $B-R$ & $218.8$ & C
& $228.5$ \\ 
\multicolumn{1}{l}{2010ay} & S12 & M16 & GH & $-$ & $297.9$ & C & $311.6$ \\ 
\multicolumn{1}{l}{2012ap} & M15 & M16 & GH & $B-V$ & $43.05$ & L & $40.37$
\\ 
\multicolumn{1}{l}{PTF10qts} & W14 & M16 & G & $g^{\prime }-i^{\prime }$ & $%
415$ & C & $428.1$ \\ 
\multicolumn{1}{l}{PTF10vgv\tablenotemark{f}} & C12 & M16 & G & $-$ & $60.3$
& C & $63.5$ \\ \hline
\end{tabular}%
\end{center}
\par
References: I00: \cite{Iwamoto00}; T06: \cite{Tomita06}; V08: \cite%
{Valenti08}; Y10: \cite{Young10}, S09: \cite{Sahu09}; P11: \cite{Pignata11};
M13: \cite{Mazzali13}; C11: \cite{Corsi11}; S12: \cite{Sanders12}; M15: \cite%
{Milisavljevic15}; W14: \cite{Walker14}; C12: \cite{Corsi12}; M16: \cite%
{Modjaz16}.\newline
\textbf{Notes.} \newline
\tablenotemark{a} None: No extinction was corrected; G: Corrected for
Galactic extinction; H: Corrected for host extinction.\newline
\tablenotemark{b} The color used to calculate bolometric magnitude,
following \cite{Lyman14}. A hyphen in this column indicates that only one
passband is available or no data in individual passbands are provided.%
\newline
\tablenotemark{c} The distance used in the referenced paper, which is
calculated according to the given distance modulus.\newline
\tablenotemark{d} The method used in this paper to calculate distance. L:
Linear distance extracted from the NASA/IPAC Extragalactic Database (NED);
C: The distance was calculated according to the latest Plank cosmological
parameters.\newline
\tablenotemark{e} The distance adopted in this paper.\newline
\tablenotemark{f} \cite{Corsi12} classified PTF10vgv as SN Ic based on its
low Si {\scriptsize II} $\lambda $6355 absorption velocities, while \cite%
{Modjaz16} reclassified it as SN Ic-BL because of its broad-lined optical
spectra. Here we follow \cite{Modjaz16}.\newline
\end{table*}

\cite{Modjaz16} listed 12 SNe Ic-BL. However, the light curve of SN 2007D is
missing and we are therefore left with 11 such events, as listed in Table %
\ref{tbl:sample}. The modeling of SN light curves usually involves the
bolometric luminosity. To construct a bolometric light curve, emission in
passbands \textit{UV} (ultraviolet), \textit{BVRI} (optical) and \textit{IR}
(UVOIR) should be integrated. It is, however, commonplace that only the
optical bands are available for the follow-up of an SN from very early times
to late times. \textit{UV} emission of an SN Ic-BL is usually strongest only
at early stages, and its contribution to the total UVOIR bolometric flux can
be more than 20\% during the first two weeks \citep{Cano11,Lyman14}, while
late-time \textit{UV} follow-up is frequently missing. \textit{IR} emission,
which is usually strong for the whole evolution stage (and can contribute as
much as 50\% of the total UVOIR bolometric flux after peak light, e.g.,
Figure 6 of \citealt{Tomita06}, Figure 14 of \citealt{Valenti08}, and Figure
7 of \citealt{Olivares15}), is only obtained for a few SNe. For this reason,
different authors usually resort to different methods to construct the
bolometric luminosity. To list some, the observations of SN 2003jd were
available only in the $BVRI$\ bands, and the contributions from $UV$\ and $%
IR $\ bands were added by assuming the same fractional contributions to the
bolometric light curve as SN 2002ap \citep{Valenti08}. The bolometric light
curve of PTF10qts was obtained by increasing the integrated fluxes by 15\%
to account for the contribution from the unavailable $UV$\ and NIR\ bands %
\citep{Walker14}. Some authors, on the other hand, decide to not include the
contribution of $UV$\ and/or NIR\ bands 
\citep{Tomita06, Sahu09, Young10,
Pignata11}.

To reduce the above uncertainty, we decide to use the method developed by 
\cite{Lyman14,Lyman16}. In this method, the color defined by two optical
bands are used to calculate the bolometric correction. In Table \ref%
{tbl:sample} we list the color we used to calculate bolometric luminosity.
In this calculation, we choose the color that has the least rms given in
Table 2 of \cite{Lyman14} and at the same time the longest time coverage in
the two passbands defining the chosen color. If these two conditions cannot
be met simultaneously, we always choose the passbands that have the longest
observational time. Such choice can minimize the errors that may be
introduced by interpolation and/or extrapolation. Sometimes data are
available only in a single passband for some time duration, e.g. the data of
SN 2007bg before 7.2 days given in Table 3 of \cite{Young10}, while these
data are crucial to constrain the fitting parameters, we set their
bolometric corrections to the same as that at the closest time.

Another uncertainty in the construction of a bolometric light curve comes
from the treatment of extinction. The Galactic extinction is well-understood
and can be handled properly using the dust maps of \cite{Schlegel98}, and as
revised by \cite{Schlafly11}. The host extinction, however, can only be
estimated for some SNe because of the poor quality of the Na {\scriptsize I}
D lines in the measured spectra 
\citep[which may be a poor proxy for the
host extinction anyways, e.g.,][]{Poznanski11}. We list the extinction
treatment in Table \ref{tbl:sample}. Even for the same SN, the determined
extinction could be different from different authors. Taking SN 2012ap as an
example, \cite{Milisavljevic15} adopted a total extinction of $E(B-V)_{%
\mathrm{total}}=0.45\unit{mag}$, while \cite{Liu15} adopted a value of $%
E(B-V)_{\mathrm{total}}=0.87\unit{mag}$.

Further uncertainty comes from the different values of the cosmological
parameters used in the literature to derive the luminosity distances to the
various SNe. For SNe 2002ap, 2009bb, 2012ap, redshift-independent methods,
e.g. Tully-Fisher measurements, were facilitated to derive the distances, as
are available on NASA/IPAC Extragalactic Database (NED). Such linear
distances are weightedly averaged, as listed in Table \ref{tbl:sample}. For
other SNe that no such linear distances are available, to minimize distance
uncertainties, we transform, according to the method described in \cite%
{Cano14}, the light curves in the literature to a common cosmology, i.e. the
latest Plank results: $H_{0}=(67.8\pm 0.9)\unit{km}\unit{s}^{-1}\unit{Mpc}%
^{-1}$, $\Omega _{m}=0.308\pm 0.012$ \citep{Ade16}.

To compare the differences in distances, we also list in Table \ref%
{tbl:sample} the distances used in the original papers. Nevertheless,
because the redshifts of the SNe studied here are small 
\citep[see Table 2
in][]{Modjaz16}, it is found that the errors in the infrared-derived
distances introduced by assuming different cosmological parameters are
small, typically $3-5\%$. \cite{Corsi12} did not give the distance modulus
of PTF10vgv in their derivation of absolute magnitudes. We digitalized their
Figure 2 and found $\mu _{\mathrm{PTF10vgv}}=33.9\unit{mag}$, based on which
the light curve transformation was performed. In summary, the largest
difference between our adopted distance and that used in the original paper
is for SN 2002ap, for which we adopt $9.22\unit{Mpc}$, rather than $7.94%
\unit{Mpc}$ in the original paper. The smallest difference is for SNe
2009bb, for which distances $\sim 40\unit{Mpc}$ have been adopted in the
relevant studies.

The photospheric velocity is another critical quantity that significantly
impacts the light curve fitting results. Different velocity indicators in
the spectra, e.g. Si {\scriptsize II} $\lambda $6355, Na {\scriptsize I} D $%
\lambda $5891, O {\scriptsize I} $\lambda $7774, Ca {\scriptsize II} $%
\lambda $8579, Fe {\scriptsize II} $\lambda $5169, usually give different
results \citep{Valenti08,Modjaz16}. This difference may be a result of the
different depth of elements in the ejecta, the degree of element mixing, and
the amount of deviation from spherical expansion. Recently \cite{Modjaz16}
developed a way of measuring velocities for all SNe Ic-BL and SNe Ic in a
homogenous way. In this paper we use the velocity data given by \cite%
{Modjaz16}, when available. Using such a homogenous data set of velocity
data reduces the bias in the resulting fitting parameters.

\begin{figure*}[tbph]
\centering\includegraphics[width=0.9\textwidth,angle=0]{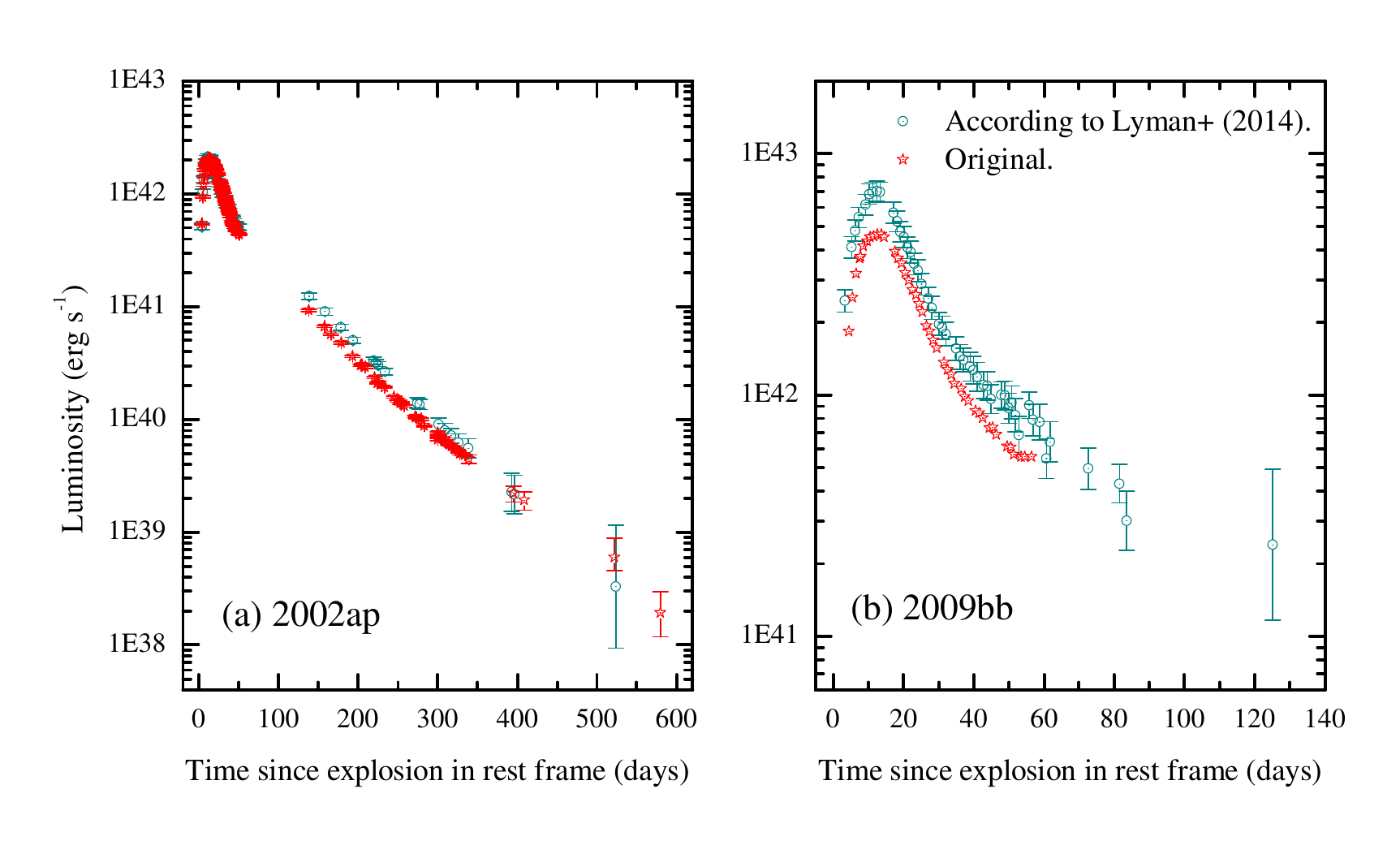}
\caption{Comparison of bolometric luminosity of two representative SNe
obtained by the method (cyan circles) of \protect\cite{Lyman14} with that
(red stars) given in the original papers. For SN 2002ap, the luminosity
given by \protect\cite{Tomita06} includes contribution from \textit{BVRI}
and \textit{IR} bands, while for SN 2009bb, the luminosity given by 
\protect\cite{Pignata11} includes contribution from \textit{BVRI} bands
only. }
\label{fig:compare-Lyman}
\end{figure*}

In principle, the above uncertainties all contribute to the errors in
bolometric luminosities. In practice, we include errors (all added in
quadrature) in bolometric corrections 
\citep[rms given in Table 2 of
][]{Lyman14} and in photometry given in the papers where the observational
data were provided.

We calculated the extinction according to \cite{Cardelli89} by assuming the
Milky Way extinction law. Cosmological expansion has been taken into account
using the following equation \citep{Hogg02, Lunnan16}%
\begin{equation}
M=m-5\log \left( D_{L}/10\unit{pc}\right) +2.5\log \left( 1+z\right) ,
\end{equation}%
where $D_{L}$\ is the luminosity distance and $z$\ is the redshift. The last
term in above equation is not a true $K$\ correction, but it is a good
approximation.

For SN PTF10vgv, only $R$-band luminosities were observed \citep{Corsi12}.
To obtain bolometric luminosities, \cite{Corsi12} assumed a bolometric
correction $M_{\mathrm{bol}}-M_{R}=-0.496\unit{mag}$ based on the early-time
photospheric temperature $T_{\mathrm{phot}}\approx 10^{4}\unit{K}$ of this
SN. We use this bolometric correction to derive the bolometric light curve
for SN PTF10vgv. Such a treatment is of course somewhat simplified because
the temperature evolves rapidly during the early expansion. Another SN for
which only $R$-band luminosities were observed is SN 2010ay \citep{Sanders12}%
. The luminosity and expansion velocity were combined to derive a
temperature of $6900\unit{K}$ at peak light. This implies a bolometric
correction $M_{\mathrm{bol}}-M_{R}=0.29\unit{mag}$, according to which the
bolometric luminosities are derived here. This treatment should not
introduce too much bias because the observation duration of this SN is
short, within $20\unit{days}$ before or after peak. For SN 1997ef, only $V$\
band data are provided by \cite{Iwamoto00}. According to the effective
temperatures ($\sim 6100\unit{K}$) given in Table 3 of \cite{Iwamoto00}, we
applied bolometric correction ($M_{\mathrm{bol}}-M_{V}=-0.05\unit{mag}$) to
SN 1997ef. We will discuss the implications of the approximation in
obtaining bolometric light curves for SNe 1997ef, 2010ay, and PTF10vgv in
Section \ref{sec:cor}.

The root mean squares of the prescription of \cite{Lyman14} are $\sim 0.06%
\unit{mag}$, while the measurement errors of the light curve range from $%
\sim 0.02\unit{mag}$\ to $\sim 0.3\unit{mag}$.\ Therefore the uncertainties
in the bolometric luminosity constructed by this method are usually
dominated by measurement errors in the two individual bands from which
bolometric corrections are calculated. The measurement errors of SNe 1997ef,
2010ay, and PTF10vgv are $0.03-0.06\unit{mag}$, $0.2-0.3\unit{mag}$, and $%
0.02-0.2\unit{mag}$, respectively. As a result, if there had been two bands
available for SNe 1997ef, 2010ay, and PTF10vgv, the uncertainties are likely
slightly larger than that depicted in Figures \ref{fig:1997ef}(a), \ref%
{fig:2010ay}(a), and \ref{fig:2003jd}(f) but dominated by measurement errors
for those points whose measurements errors are large. Given this fact, for
simplicity, we adopt the errors in an individual band as the errors of
bolometric luminosity for SNe 1997ef, 2010ay, and PTF10vgv.\footnote{%
However, given the several sources of uncertainty in the bolometric
correction derived from a single filter, we estimate that the minimum error
in the latter must be at least 20\%, if not larger.}

In Figure \ref{fig:compare-Lyman} we compare the luminosity data provided by
the original papers and that calculated according to \cite{Lyman14} for two
representative SNe. We call the luminosity of these two SNe `representative'
because the luminosity of SN 2002ap given by the original paper includes the
contribution from $BVRI$\ and $IR$\ bands, while the luminosity of SN 2009bb
given by the original paper includes contribution only from $BVRI$\ bands.
Another reason we choose these two SNe is that their luminosity is
integrated according to observational data, while the luminosity of some
other SNe are calculated in the original papers by assuming some
contribution from unavailable bands (frequently the $IR$\ band).

In the comparison in Figure \ref{fig:compare-Lyman} the data given in the
original papers are transformed to the distances given in Table \ref%
{tbl:sample}. From this figure it is evident that the method of \cite%
{Lyman14} is accurate for the first $\sim 80\unit{days}$,\ since the
bolometric correction is calculated according to the luminosity data in this
time period. Fortunately, most of the luminosity data in our sample have a
time coverage that is not much longer than $\sim 80\unit{days}$. The data
with $t\gtrsim 80\unit{days}$\ are enough to constrain most of the model
parameters. Figure \ref{fig:compare-Lyman}(a) shows that the contribution
from the unavailable $UV$\ band is small for SN 2002ap even at very early
stages\footnote{%
The $UV$ flux observed on 2002 February 3 (4 days before $V$-band maximum)
by \emph{XMM-Neutron} contributes only $\sim 4\%$ \citep{Mazzali02}.}, while
Figure \ref{fig:compare-Lyman}(b) shows that the contribution from the
unavailable $UV$\ and $IR$\ bands cannot be ignored.

\section{Fitting result}

\label{sec:result}

As explained in \cite{WangWang16,WangYu17}, the model we have adopted is
formulated by eight parameters. Although the model is dubbed a
\textquotedblleft magnetar model\textquotedblright , it also includes a $%
^{56}$Ni component. As a consequence, the model includes the usual
parameters, the ejecta mass $M_{\mathrm{ej}}$, $^{56}$Ni mass $M_{\mathrm{Ni}%
}$, grey optical opacity $\kappa $, initial expansion velocity $v_{\mathrm{sc%
}0}$, and opacity to $^{56}$Ni decay photons $\kappa _{\gamma ,\mathrm{Ni}}$%
. In addition, the model includes magnetar parameters, the dipole magnetic
field $B_{p}$, initial rotation period $P_{0}$ and opacity $\kappa _{\gamma ,%
\mathrm{mag}}$ to account for the leakage \citep{Chen15,WangWang15} of high
energy photons \citep{Murase15,WangDai16} from magnetars. Here the subscript
\textquotedblleft p" in $B_{p}$\ means the dipole field at the pole of the
star \citep{Shapiro83}. For the grey optical opacity $\kappa $ we take the
fiducial value $\kappa =0.1\unit{cm}^{2}\unit{g}^{-1}$, as used in previous
investigations \citep{WangHan16,
WangYu17}. We also include the unknown explosion time $T_{\mathrm{start}}$
of the SN in the MCMC code. In what follows, we use the name
\textquotedblleft magnetar model" to indicate the magnetar+$^{56}$Ni model,
except specifically mentioned otherwise.

\begin{sidewaystable}[tbph]
\caption{Best-fitting parameters of our SNe Ic-BL sample}
\begin{center}
\begin{tabular}{ccccccccc|ccc}
\hline\hline
SN & $M_{\mathrm{ej}}$ & $M_{\mathrm{Ni}}$ & $B_{p}$ & $P_{0}$ & $v_{\mathrm{%
sc}0}$ & $\kappa _{\gamma ,\mathrm{Ni}}$ & $\kappa _{\gamma ,\mathrm{mag}}$
& $T_{\mathrm{start}}$ & $T_{\mathrm{Neb}}$ & Constraints on $T_{\mathrm{Neb}%
}$\tablenotemark{a} & $E_{K0}$ \\ 
\multicolumn{1}{l}{} & $\left( M_{\odot }\right) $ & $\left( M_{\odot
}\right) $ & $\left( 10^{15}\unit{G}\right) $ & $\left( \unit{ms}\right) $ & 
$\left( \unit{km}\unit{s}^{-1}\right) $ & $\left( \unit{cm}^{2}\unit{g}%
^{-1}\right) $ & $\left( \unit{cm}^{2}\unit{g}^{-1}\right) $ & $\left( \unit{%
days}\right) $ & $\left( \unit{days}\right) $ & $\left( \unit{days}\right) $
& $10^{51}\unit{erg}$ \\ \hline
\multicolumn{1}{l}{1997ef} & \multicolumn{1}{l}{$3.3_{-0.19}^{+0.21}$} & $%
0.058\pm 0.001$ & $11.4_{-0.5}^{+0.4}$ & $5.2_{-0.3}^{+1.3}$ & $%
1900_{-1000}^{+2600}$ & $0.12_{-0.02}^{+0.03}$ & $-$ & $-16.1_{-0.5}^{+0.4}$
& $153$ & $\left( 61.3,119.8\right) $ & $0.07$ \\ 
\multicolumn{1}{l}{2002ap} & \multicolumn{1}{l}{$1.7\pm 0.2$} & $0.049\pm
0.001$ & $19.5\pm 2$ & $12\pm 4$ & $10500_{-1000}^{+700}$ & $0.20\pm 0.02$ & 
$8.8\pm 6$ & $1.9\pm 0.1$ & $71$ & $\left( 44,81\right) $ & $1.13$ \\ 
\multicolumn{1}{l}{2003jd} & $2.8\pm 0.3$ & $0.05\pm 0.02$ & $2.5\pm 0.3$ & $%
18_{-1.6}^{+1.2}$ & $12490\pm 500$ & $0.3_{-0.2}^{+0.3}$ & $%
1.75_{-1.1}^{+1.5}$ & $-14.9\pm 1$ & $92$ & $\left( 62.5,81.2\right) $ & $%
2.64$ \\ 
\multicolumn{1}{l}{2007bg} & $1.5\pm 0.2$ & $0.03\pm 0.004$ & $10.5_{-3}^{+5}
$ & $20_{-13}^{+11}$ & $11900_{-2100}^{+1200}$ & $0.4_{-0.3}^{+1.2}$ & $-$ & 
$-10\pm 2$ & $77$ & $\left( 34.7,67.7\right) $ & $1.3$ \\ 
\multicolumn{1}{l}{2007ru} & $5.6\pm 0.5$ & $0.078\pm 0.006$ & $%
6.2_{-0.2}^{+0.3}$ & $1.6\pm 0.1$ & $\sim 0$ & $0.19_{-0.03}^{+0.05}$ & $-$
& $0.8_{-0.6}^{+0.9}$ & $86$ & $\left( 70,200\right) $ & $\sim 0$ \\ 
\multicolumn{1}{l}{2009bb} & $2.2\pm 0.07$ & $0.025\pm 0.01$ & $2.9\pm 0.3$
& $28.6\pm 0.6$ & $19300_{-700}^{+600}$ & $0.4_{-0.35}^{+2.8}$ & $%
2.3_{-1.6}^{+1.8}$ & $-12\pm 0.4$ & $48$ & $\left( 55,295\right) $ & $4.90$
\\ 
\multicolumn{1}{l}{2010ah} & $2.58_{-1.6}^{+2.6}$ & $0.14_{-0.03}^{+0.02}$ & 
$17.5_{-7.9}^{+20}$ & $4.2_{-2.7}^{+35}$ & $14100_{-9700}^{+10000}$ & $%
\gtrsim 0.06$ & $-$ & $-0.7\pm 1$ & $69$ & $>14.5$ & $3.1$ \\ 
\multicolumn{1}{l}{2010ay} & $6.7_{-1}^{+1.8}$ & $-$ & $0.8\pm 0.1$ & $11\pm
1$ & $24900_{-2100}^{+300}$ & $-$ & $-$ & $2.0_{-0.4}^{+0.3}$ & $55$ & $>44.1
$ & $33.4$ \\ 
\multicolumn{1}{l}{2012ap} & $2.3_{-0.8}^{+1.7}$ & $-$ & $3.1_{-0.5}^{+1.6}$
& $40_{-11}^{+3}$ & $14219_{-900}^{+1000}$ & $-$ & $-$ & $-0.6\pm 1$ & $77$
& $\left( 38.5,230.5\right) $ & $2.72$ \\ 
\multicolumn{1}{l}{PTF10qts} & $1.9_{-0.4}^{+0.6}$ & $0.28\pm 0.06$ & $%
9.9_{-6.6}^{+12}$ & $26_{-19}^{+64}$ & $21307_{-2500}^{+2400}$ & $%
0.3_{-0.1}^{+1.5}$ & $-$ & $-10.5_{-1.9}^{+2.7}$ & $39$ & $\left(
38.8,230.4\right) $ & $5.1$ \\ 
\multicolumn{1}{l}{PTF10vgv} & $0.7_{-0.07}^{+0.08}$ & $0.059\pm 0.003$ & $%
1.7\pm 0.2$ & $28.8_{-0.7}^{+0.5}$ & $10000_{-600}^{+700}$ & $0.19\pm 0.03$
& $0.013\pm 0.001$ & $6.9\pm 0.04$ & $43$ & $\left( 46.2,82.6\right) $ & $%
0.42$ \\ \hline
\end{tabular}%
\end{center}
\par
\textbf{Notes.} \newline
All times in this table are in rest frame.\newline
In these fits, we fixed $\kappa =0.1\unit{cm}^{2}\unit{g}^{-1}$. \newline
A hyphen indicates that this quantity cannot be constrained effectively.%
\newline
The data on the left of the vertical line are fitting parameters, while the
data on the right are derived values. The MCMC code does not calculate the
errors of these derived values. \newline
The references for the spectra are the same as in Table \ref{tbl:sample}.
\label{tbl:para}
\end{sidewaystable}

The magnetar model proposed by \cite{WangWang16} traces the photospheric
recession and therefore the emission from the photosphere and those material
outside of the photospheric radius (hereafter referred to as nebular
component, or nebula for short) can be isolated. We examined the spectra of
SNe Ic-BL and try to figure out if the nebular emission is helpful in
determining the appearance of nebular features in the spectra. It turns out
that when the nebula emitted about $16\%$ of the total emission, nebular
features (e.g., forbidden lines) could begin to emerge in the SN spectra. If
we assume that an SN begins to transition into nebular phase when the nebula
radiates this percentage of emission, we can obtain the time $T_{\mathrm{Neb}%
}$ (since explosion in rest frame) by fitting the early-time light curve, as
listed in Table \ref{tbl:para}.\footnote{%
This approach is desirable as it is sometimes difficult to identify an
eruption as a SN or a tidal disruption event %
\citep[TDE;][]{Brown16,Dong16,Leloudas16}. The indication of the early
nebular phase is helpful to confirm the identity of a SN because a TDE does
not have a nebular phase. We therefore encourage the modeling of the
early-time light curve of a transient to give an estimate of the epoch at
which nebular features may appear (if it is a SN) to help constrain the
nature of the transient.} In the magnetar model, the early peak of the light
curve of an SN is caused by the spin-down of the magnetar. Consequently the $%
^{56}$Ni mass can be ignored for such early-time modeling.

\begin{figure*}[tbph]
\centering\includegraphics[width=1%
\textwidth,angle=0]{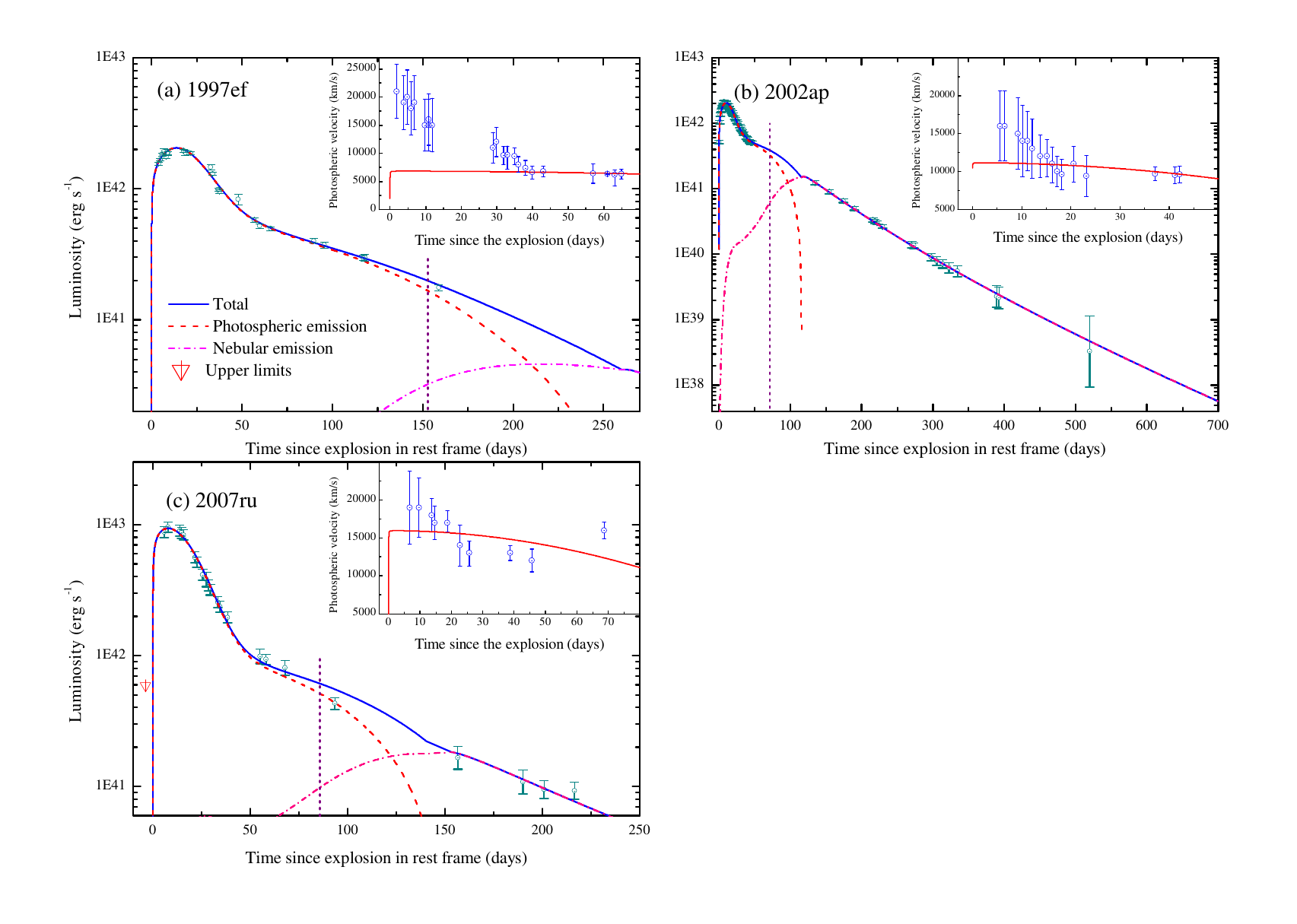}
\caption{The magnetar+$^{56}$Ni model: best-fitting light curves (solid
lines) of SNe 1997ef, 2002ap, 2007ru. The dashed and dot-dashed lines arise
from photospheric and nebular emission, respectively. The vertical dotted
lines mark the time when nebular emission lines becomes significant. The
insets show the fit (red solid lines) to the evolution of photospheric
velocities.}
\label{fig:1997ef}
\end{figure*}

To determine the $^{56}$Ni mass, it is necessary for the light curve to be
observed at least for $\sim 110\unit{days}$. We divide the observed SNe
Ic-BL light curves into two classes: those with an observational duration $%
t\gtrsim 100\unit{days}$ (class I) and with $t\lesssim 100\unit{days}$
(class II). The reason for the choice of $100\unit{days}$ as the dividing
boundary is because the lifetime of $^{56}$Co is $\sim 110\unit{days}$. If
observational duration is longer than $100\unit{days}$, the mass of $^{56}$%
Ni can be constrained. In this case we allowed the $^{56}$Ni mass to be a
free parameter. In the opposite case, the $^{56}$Ni mass cannot be
constrained and the only parameters that can be constrained are the magnetar
parameters because it is found that in the magnetar model the early peak of
the light curve can be attributed to magnetar spin-down 
\citep{WangHan16,
WangYu17}.

\begin{figure*}[tbph]
\centering\includegraphics[width=1%
\textwidth,angle=0]{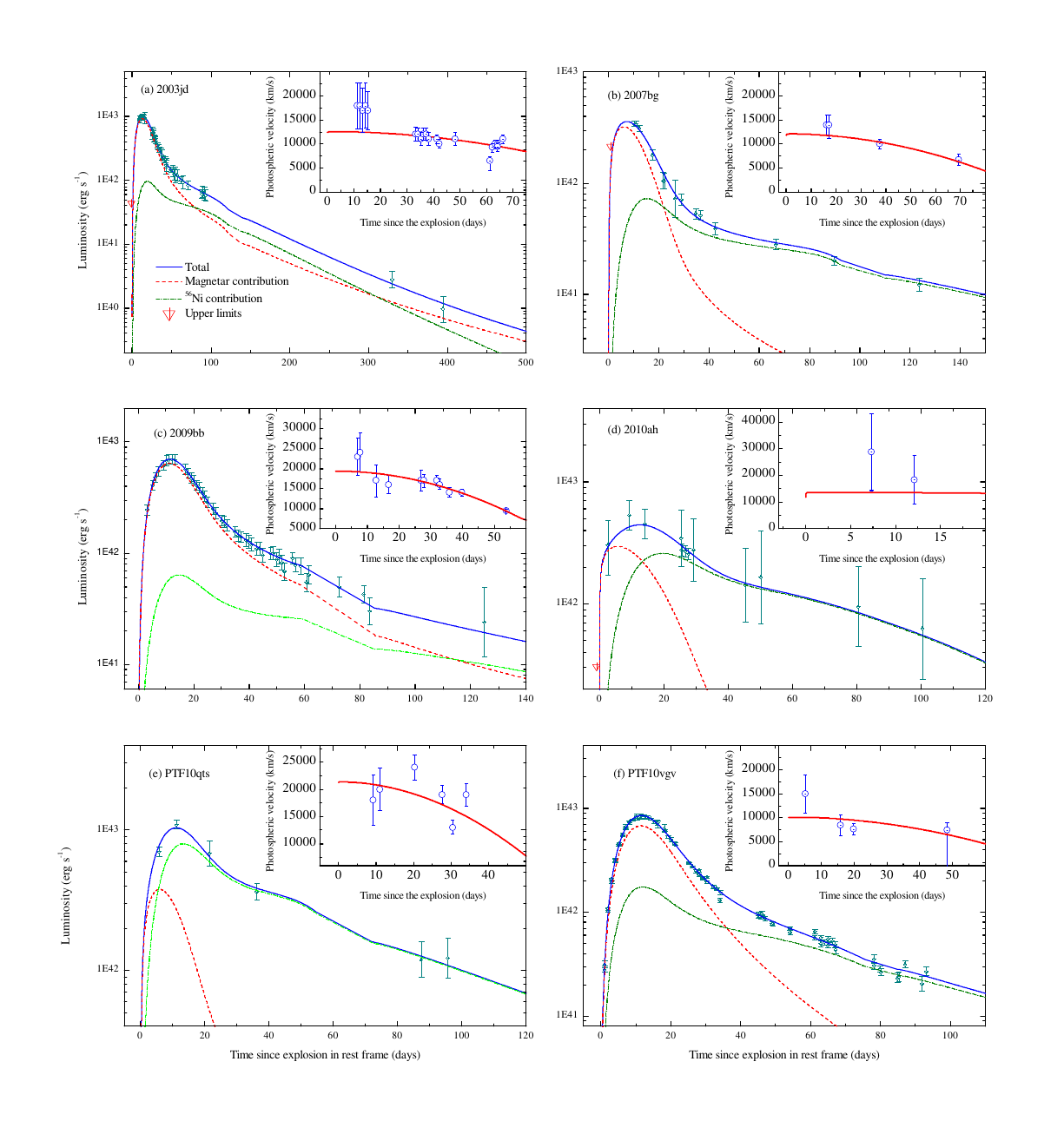}
\caption{The magnetar+$^{56}$Ni model: best-fitting light curves (solid
lines) of SNe 2003jd, 2007bg, 2009bb, 2010ah, PTF10qts, and PTF10vgv. The
dashed lines are the contribution from magnetar, while the dot-dashed lines
show the contribution from $^{56}$Ni. The insets show the fit (red solid
lines) to the evolution of photospheric velocities.}
\label{fig:2003jd}
\end{figure*}

We find there are nine SNe that belong to class I, while the remaining two
SNe fall in class II. Among the SNe in class I, three SNe, 1997ef, 2002ap,
and 2007ru, were studied previously with our magnetar model %
\citep{WangHan16, WangYu17}. SN 2002ap was investigated using an MCMC code %
\citep{WangYu17}, while SNe 1997ef and 2007ru were studied via manual
fitting \citep{WangHan16}. We included them here to test the sensitivity of
fitting parameters to the adoption of different photospheric velocities
because the velocities used here \citep[the values
given by][]{Modjaz16} are different from previous studies 
\citep{WangHan16,
WangYu17} where we used the velocities provided in the original papers. In
addition, doing so will give unbiased statistical results.

\begin{figure*}[tbph]
\centering\includegraphics[width=1\textwidth,angle=0]{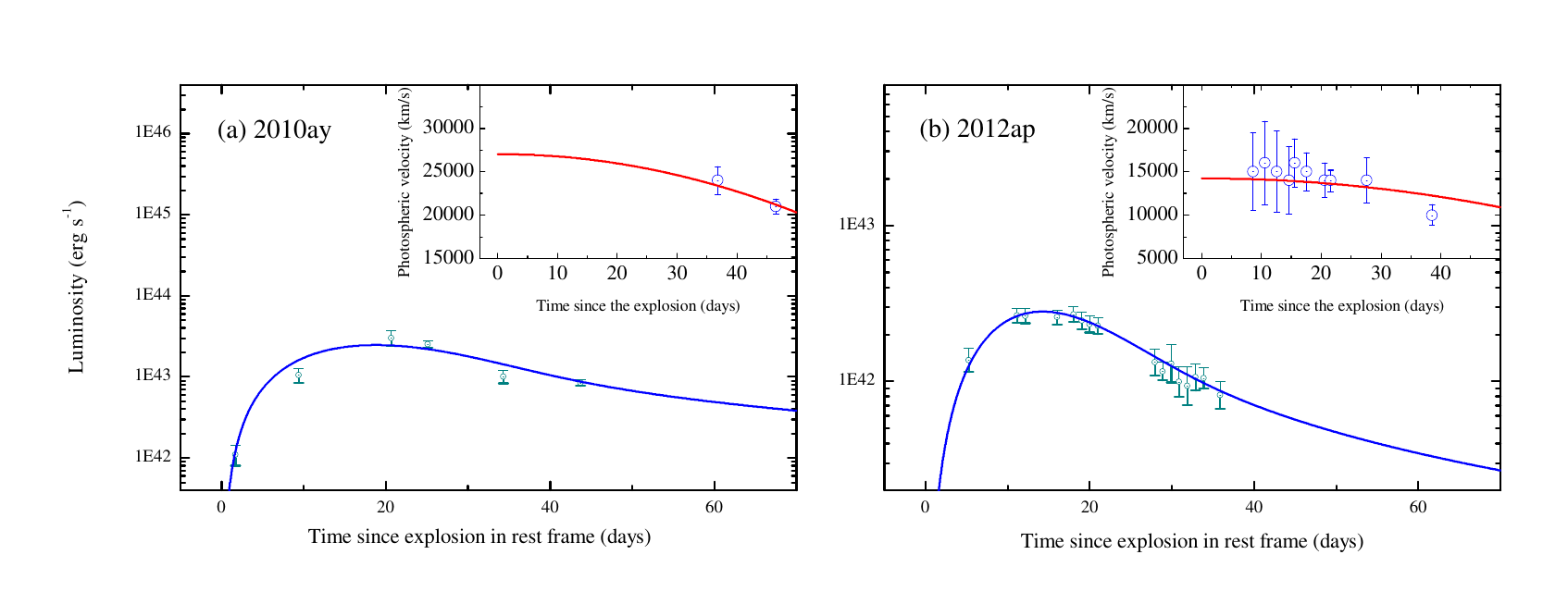}
\caption{Light curves of SNe 2010ay, and 2012ap fitted by a pure-magnetar
(without $^{56}$Ni) model. The insets show the fit (red solid lines) to the
evolution of photospheric velocities.}
\label{fig:2010ay}
\end{figure*}

The newly fitted light curves of the three previously studied SNe 1997ef,
2002ap, and 2007ru are shown in Figure \ref{fig:1997ef}, where emission from
the photosphere and nebula are shown as dashed and dot-dashed lines,
respectively. The vertical dotted lines in this figure mark the epochs when
nebular emission contributes 16\% of the total emission, which we assume to
be the time when nebular emission features, e.g. forbidden emission lines,
begin to appear. The remaining six SNe in class I are shown in Figure \ref%
{fig:2003jd}, where we show the contribution from the magnetar and $^{56}$Ni
as dashed and dot-dashed lines, respectively. In Figure \ref{fig:2010ay} the
light curves were fitted with a pure magnetar model (without $^{56}$Ni
contribution) because the $^{56}$Ni masses of the these two SNe in class II
cannot be determined. The best-fitting values are given in Table \ref%
{tbl:para}. Because the contribution of magnetar and $^{56}$Ni to the total
emission for the three SNe depicted in Figure \ref{fig:1997ef} were shown
previously \citep{WangHan16, WangYu17}, we will not show them in this paper,
as we do in Figure \ref{fig:2003jd}.

\begin{figure}[tbph]
\centering\includegraphics[width=0.48\textwidth,angle=0]{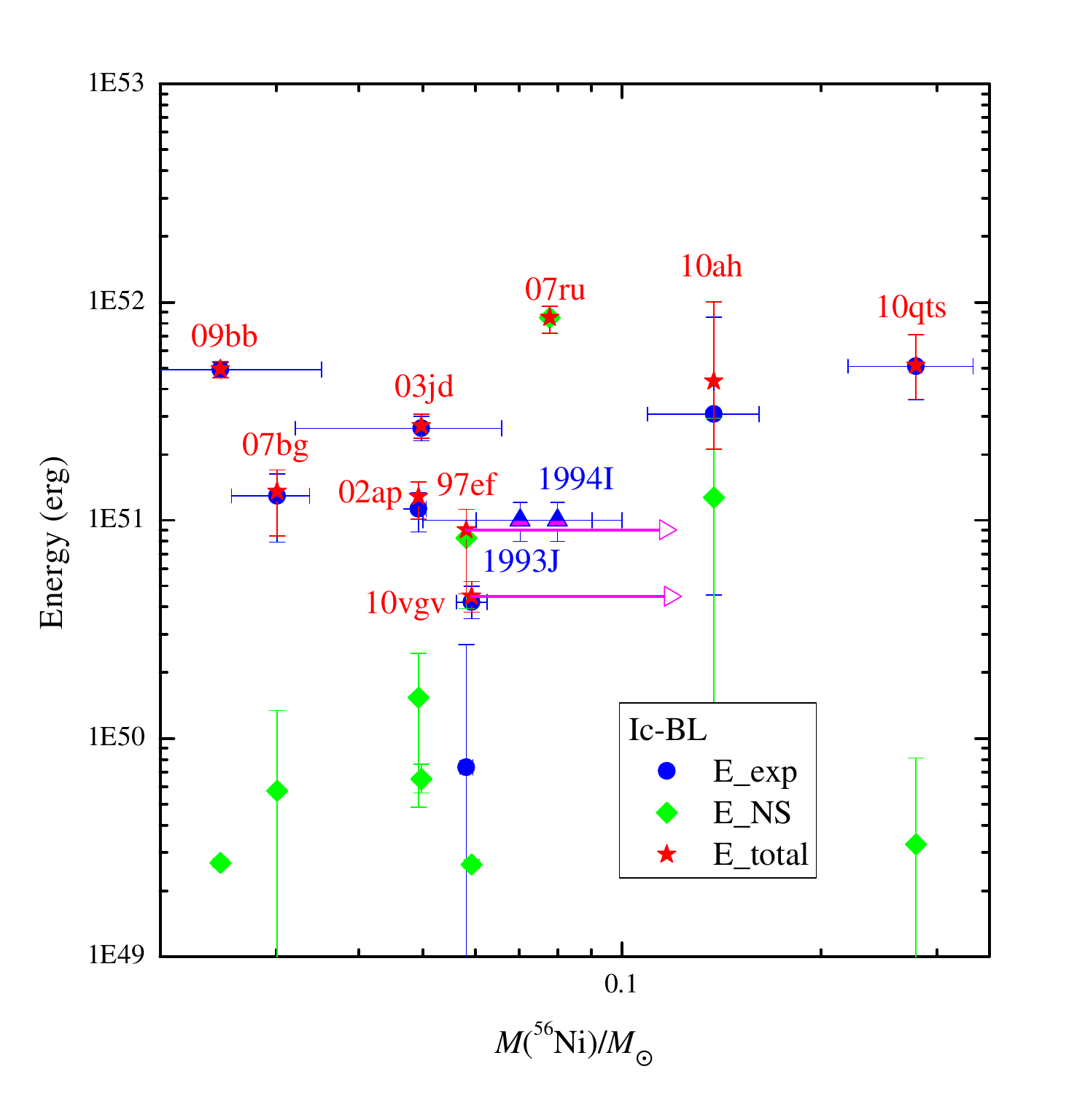}
\caption{$^{56}$Ni mass versus explosion energy, neutron star rotational
energy, and the sum of these two energies. The ordinary SNe 1993J (IIb) and
1994I (Ic) are plotted as triangles. For clarity, only one $^{56}$Ni error
bar is shown for each SN. For SNe 1997ef and PTF10vgv whose bolometric light
curves are constructed from single bands, we show the uncertainties in $^{56}
$Ni mass as horizontal magenta lines with rightward arrows.}
\label{fig:Ni-E}
\end{figure}

For SNe 2007bg, 2010ah, and PTF10qts in Figure \ref{fig:2003jd}, the
magnetar contribution dies away rapidly, which is quite different from the
light curves given by e.g. \cite{Kasen10}, where the light curves tend to
flatten at late time. The decline rate is also faster than the light curves
where the gamma-ray leakage has been taken into account 
\citep{Chen15,
WangWang15}. This rapid decline of magnetar contribution is due to the rapid
spin-down of the magnetar powering the SN Ic-BL. At very late times, the
magnetar contribution will eventually flatten, as can be seen from the
dashed lines in Figures 1 and 2 of \cite{WangYu17}. This rapid spin-down is
why the magnetar can convert almost all of its rotational energy to the
kinetic energy of ejecta of SNe Ic-BL and why the contribution of $^{56}$Ni
is necessary for SNe Ic-BL in the magnetar model.

\begin{figure}[tbph]
\centering\includegraphics[width=0.48\textwidth,angle=0]{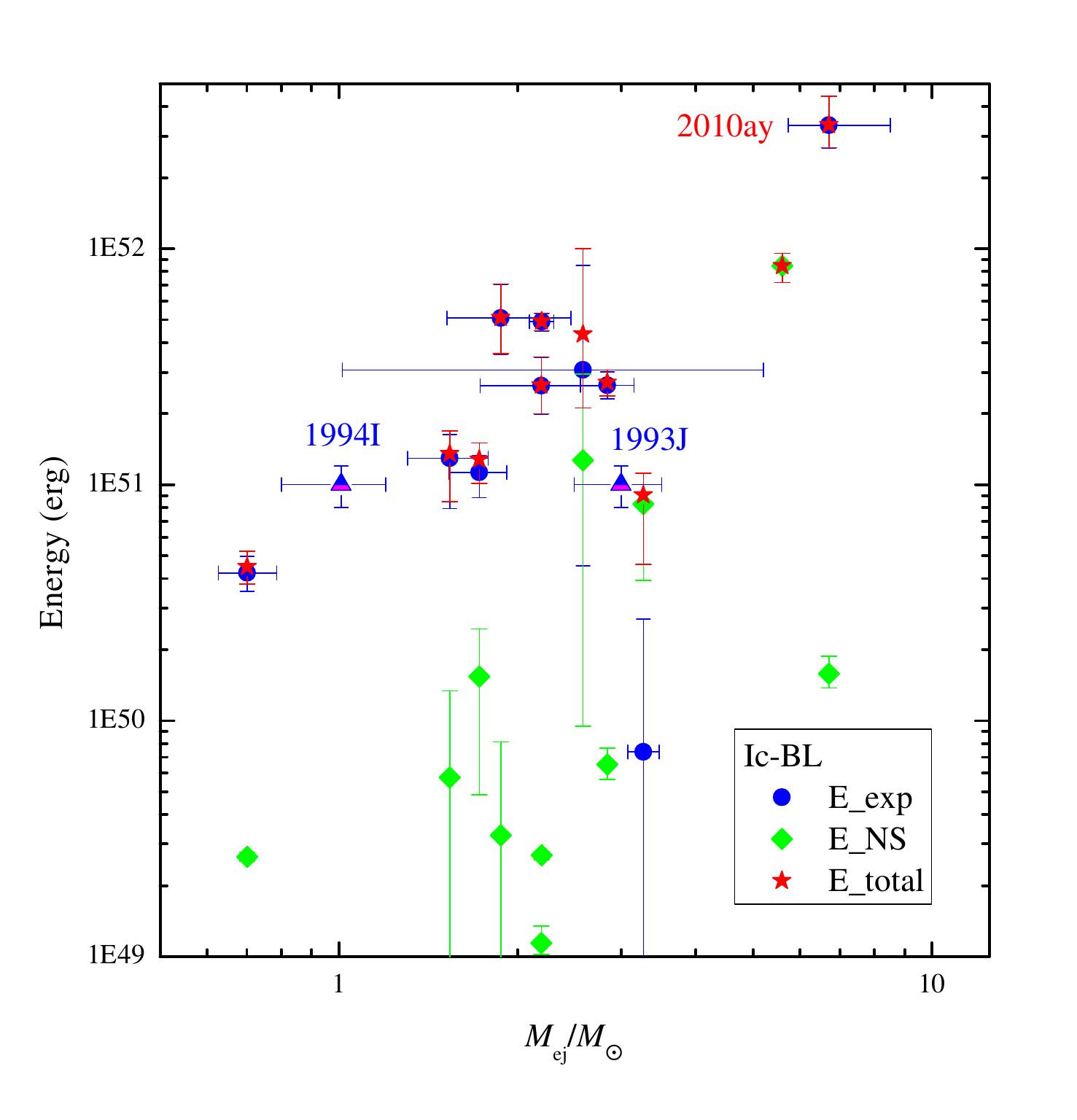}
\caption{Ejecta mass versus explosion energy, neutron star rotational
energy, and the sum of these two energies. The ordinary SNe 1993J (IIb) and
1994I (Ic) are plotted as triangles. For clarity, only one ejecta mass error
bar is shown for each SN.}
\label{fig:Mej-E}
\end{figure}

The MCMC code can only be run for those SNe for which observational errors
are given. For the velocity data given by \cite{Modjaz16}, we adopted the
errors given in their paper. \cite{Modjaz16} did not provide the velocity
data for SN 2010ah, for which, following \cite{WangYu17}, we set the
velocity errors to be half of the measured values to account for the large
differences given by different velocity measurements 
\citep[see
e.g.,][]{Valenti08}.

For some SNe, e.g. SNe 2007bg and 2007ru, the missing or sparse data
coverage before peak luminosity makes the upper limits before discovery
indispensable for obtaining reliable results, see Figures \ref{fig:1997ef}%
(c) and \ref{fig:2003jd}(b). Sometimes the model light curves almost pass
through the bolometric limits, e.g. SNe 2003jd and 2007bg in Figure \ref%
{fig:2003jd}. We looked into the data and found that the bolometric limit of
SN 2003jd is $0.6\unit{days}$\ earlier than the explosion date, while the
bolometric limit of SN 2007bg is $1.0\unit{days}$\ earlier than the time
when the model light curve reaches the same bolometric luminosity as the
upper limit.

The best-fitting parameters listed in Table \ref{tbl:para} are generally
similar to what we found before for SNe 1997ef, 1998bw, 2002ap, and 2007ru 
\citep{WangHan16,
WangYu17}, where we had a detailed discussion on the reasonability of the
determined parameters such as $M_{\mathrm{ej}}$, $M_{\mathrm{Ni}}$, $B_{p}$, 
$P_{0}$, $\kappa _{\gamma ,\mathrm{mag}}$. We also discussed the possible
reasons for a larger value of $\kappa _{\gamma ,\mathrm{Ni}}$ than the
standard value $\sim 0.027\unit{cm}^{2}\unit{g}^{-1}$.

\begin{figure}[tbph]
\centering\includegraphics[width=0.48\textwidth,angle=0]{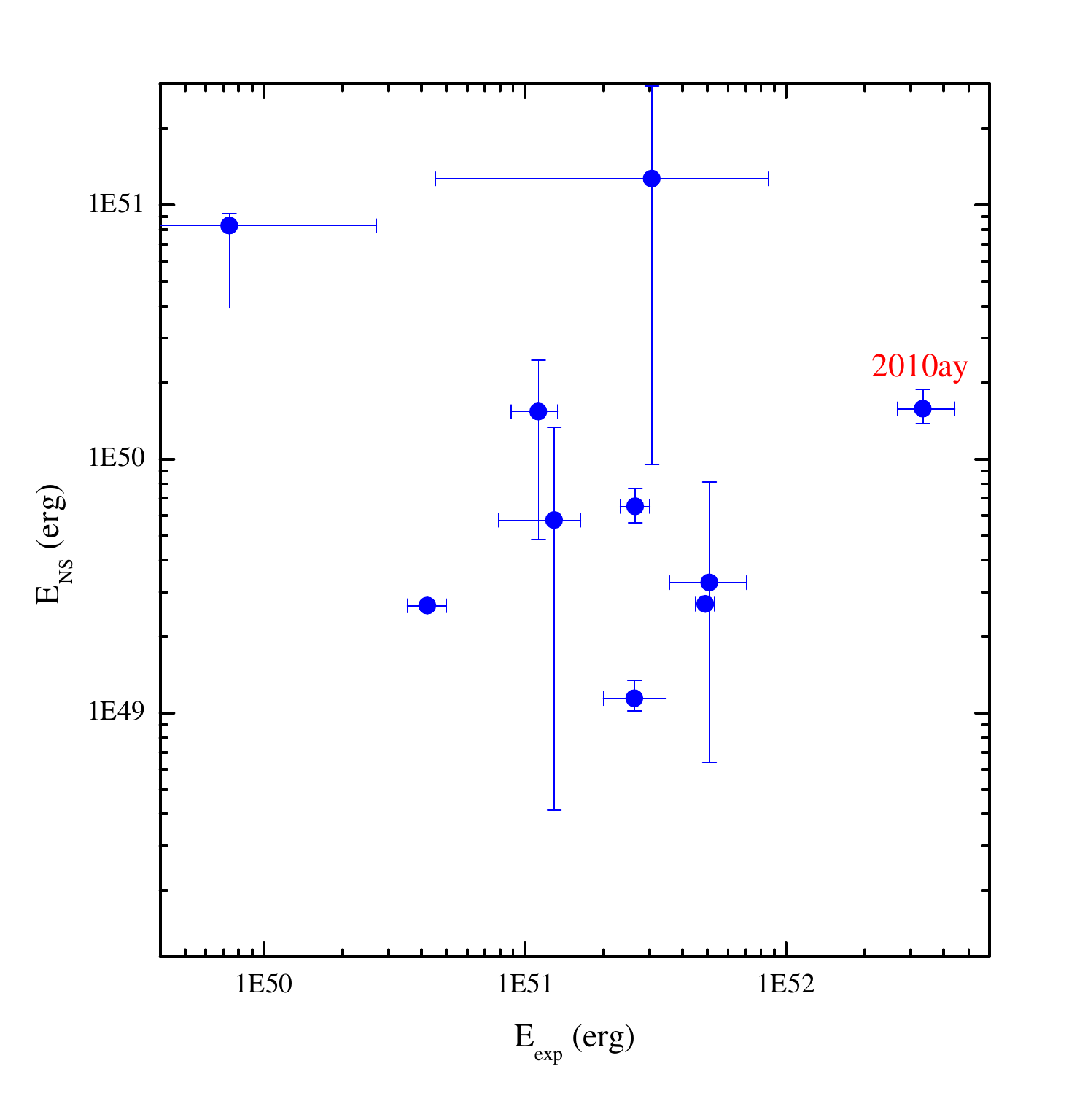}
\caption{Explosion energy versus neutron star rotational energy.}
\label{fig:Eexp-Ens}
\end{figure}

From Table \ref{tbl:para} it is clear that usually the opacity to magnetar
high-energy photons $\kappa _{\gamma ,\mathrm{mag}}$ can only be determined
for SNe, e.g. 2002ap and 2003jd, observed to late stages ($t\gtrsim 300\unit{%
days}$) because only at such late stages (except for the early peak) does
the magnetar contribution dominate the $^{56}$Ni contribution. Table \ref%
{tbl:para} indicates that $\kappa _{\gamma ,\mathrm{mag}}$ is also
constrained for SNe 2009bb and PTF10vgv, despite their short observation
duration. For PTF10vgv, the given value is favored because the ejecta mass
is small and high-energy radiation from the magnetar will leak even at early
stages. For SN 2009bb, the given value is caused by the significant
contribution of magnetar to the light curve even at late stages.

\begin{figure}[tbph]
\centering\includegraphics[width=0.48\textwidth,angle=0]{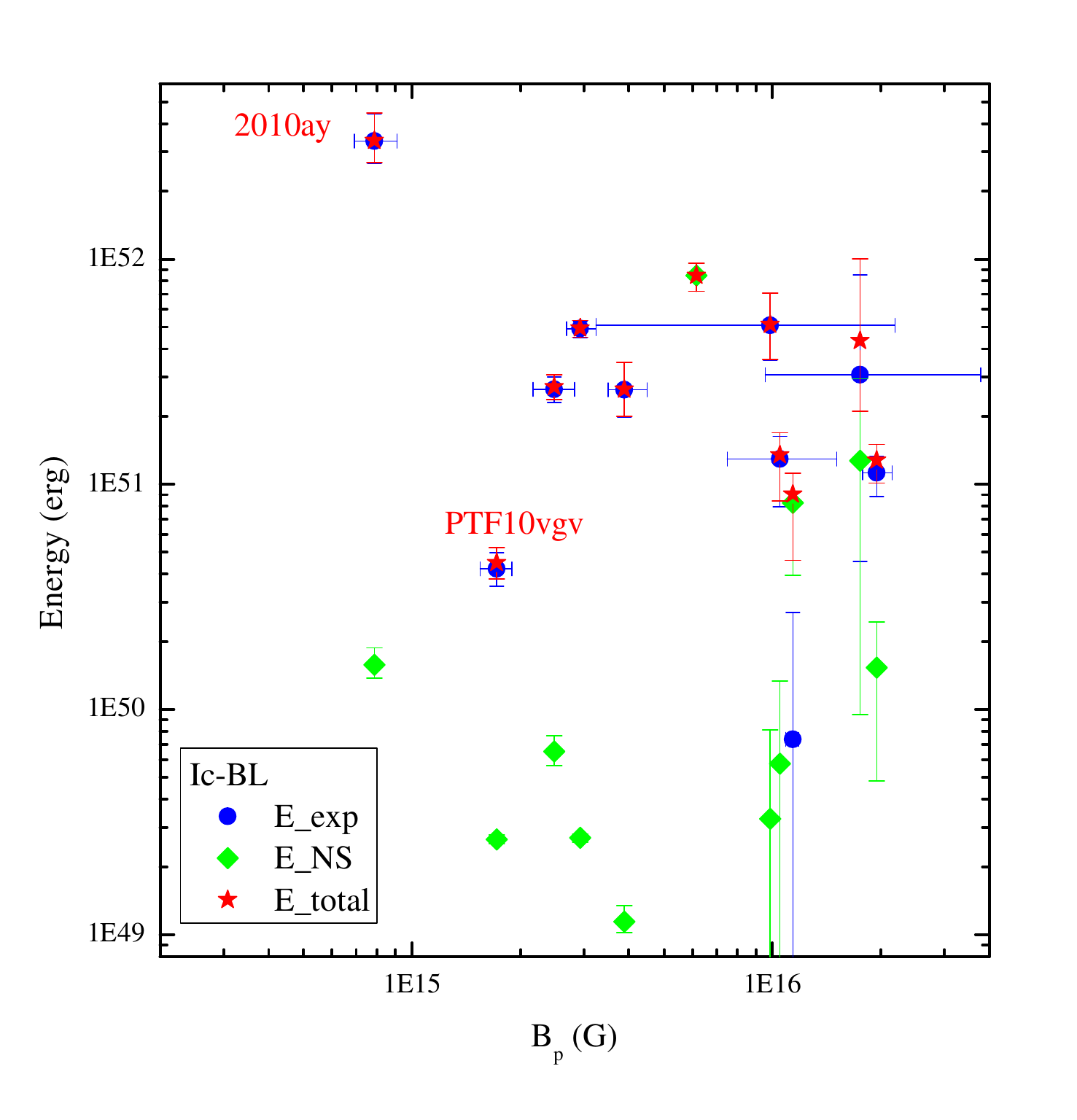}
\caption{$B_{p}$ versus explosion energy, neutron star rotational energy,
and the sum of these two energies. For clarity, only one $B_{p}$ error bar
is shown for each SN.}
\label{fig:Bp-E}
\end{figure}

Given the fitting results in Table \ref{tbl:para}, some correlations between
different parameters can be examined. We show the correlations of energy
versus $^{56}$Ni mass, energy versus ejecta mass, explosion energy versus
neutron star rotational energy, energy versus dipole magnetic field of the
magnetar in Figures \ref{fig:Ni-E}-\ref{fig:Bp-E}, respectively. In the
magnetar model, three forms of energy are considered here, i.e. the initial
explosion energy $E_{\exp }$, the neutron star's rotational energy $E_{%
\mathrm{NS}}$ and the sum of these two energies $E_{\mathrm{total}}$. In the
usual magnetar model that does not take into account the acceleration of the
ejecta by the spinning-down magnetar, the kinetic energy of the ejecta is
just the initial explosion energy. In our adopted model, the kinetic energy
is no longer a constant. Instead, it evolves from its initial value, i.e.
the initial explosion energy, according to the energy injection of the
magnetar. The evolution of kinetic energy can be clearly appreciated by
inspecting the rapid rise of the photospheric velocities at early times,
e.g. SNe 1997ef, 2007ru in Figure \ref{fig:1997ef}. This is why the reported
(initial) velocities of these two SNe are significantly smaller than the
maximum values attained in Figure \ref{fig:1997ef}.

\section{Discussion}

\label{sec:dis}

\subsection{General implications}

\label{sec:gen-imp}

The high fitting quality both for luminosities and velocities can be
appreciated from Figures \ref{fig:1997ef}-\ref{fig:2010ay}. The only
exception may be the velocity fitting result of SN 1997ef and to a lesser
extent SN~2003jd. For SN 1997ef, the early-time velocities cannot be fitted
because in our one-zone model the velocity increases rapidly at very early
times (acceleration phase) and then declines progressively faster because of
the photospheric recession. Inspection of the velocity data of SN 1997ef
indicates that the velocity evolution is flat during the period 40-65 days,
which implies that the earlier time velocity should also be flat in our
model. A possible way of getting better fitting results may be the
introduction of a fast-moving shell in the ejecta, which makes the
photosphere begin to recede at early times. After this fast shell becomes
transparent, the inner compact component slows down the recession of the
photosphere, resulting in the later-time flat evolution of the velocity.
This hypothesis is also supported by the earlier appearance of nebular
spectrum than our prediction (see Section \ref{sec:T_Neb} for more
discussion). The model velocity of SN 2010ah (Figure \ref{fig:2003jd}) is
also somewhat lower compared with the data. This is caused by the large
observational errors which make the MCMC code difficult to differentiate
between different fitting parameters.

The effect of adopting different velocity data on the derived parameters can
be appreciated by comparing the values of the fitting parameters of SN
1997ef given in Table 1 of \cite{WangHan16} and Table \ref{tbl:para}. It
turns out that the values of $M_{\mathrm{Ni}}$, $B_{p}$, $\kappa _{\gamma ,%
\mathrm{Ni}}$, $\kappa _{\gamma ,\mathrm{mag}}$, and $T_{\mathrm{start}}$
are insensitive to the expansion velocity, while $M_{\mathrm{ej}}$, $P_{0}$,
and $v_{\mathrm{sc}0}$ \emph{are} sensitive to the expansion velocity. This
is because the former group of parameters are determined by the light curve
slope ($\kappa _{\gamma ,\mathrm{Ni}}$, $\kappa _{\gamma ,\mathrm{mag}}$,
and $T_{\mathrm{start}}$) or luminosity ($M_{\mathrm{Ni}}$\ and $B_{p}$),
while the latter group of parameters are determined by the diffusion time
scale of the SN \citep{Arnett82, Arnett17} because $P_{0}$\ and $v_{\mathrm{%
sc}0}$\ affect the expansion velocity.

From Table \ref{tbl:para} it is evident that in the magnetar (plus $^{56}$%
Ni) model, the initial SN explosion energies are usually smaller than $\sim
2.5\times 10^{51}\unit{erg}$, i.e. the theoretical upper limit of explosion
energy triggered by neutrino heating \citep{Janka16}. There is one
exception, SN 2010ay, which has explosion energy $\gtrsim 10^{52}\unit{erg}$%
. We note that the light curve and velocity evolution of SN 2010ay are
poorly sampled and it is possible to attribute a fraction of the energy to
magnetar by tuning up the rotational energy of the magnetar. We conclude
that the explosion energy of all well-observed SNe Ic-BL can be explained by
neutrino heating.

Table \ref{tbl:para} shows that the $^{56}$Ni masses in this sample of SNe
Ic-BL are usually smaller than $0.1M_{\odot }$. The only two values $%
0.28M_{\odot }$\ and $0.14M_{\odot }$\ that are above $0.1M_{\odot }$\ are
for PTF10qts and SN 2010ah, respectively. We note that the observational
data of these two SNe are of poorest quality, except SNe 2010ay and 2012ap
whose $^{56}$Ni masses are not determined. The sparse luminosity data and
large observational errors of these two SNe indicate that the derived values
of $^{56}$Ni mass should not be taken seriously. This implies that the $%
^{56} $Ni masses of SNe Ic-BL have an upper limit $0.2M_{\odot }$, i.e. the
maximal amount of $^{56}$Ni that can be synthesized by the spin-down of a
magnetar \citep{Nishimura15, Suwa15}.

Table \ref{tbl:para} also shows that the fitting parameters of the two
relativistic SNe, 2009bb and 2012ap, are typical among this SNe Ic-BL
sample. It is therefore unlikely to acquire more clue on the explosion
mechanism of relativistic SNe solely from such fitting parameters, if the
magnetar model is the right model for such SNe. A thorough comparison
between SNe Ic-BL and those associated with GRBs is required to get more
clue.

PTF10vgv is peculiar because of its low absorption velocities (typical of
ordinary SNe Ic) and broad-lined optical spectra (typical of SNe Ic-BL). It
has the lowest ejecta mass, $0.7M_{\odot }$, in the SNe Ic-BL sample (see
Table \ref{tbl:para}). Its opacity to magnetar photons, $\kappa _{\gamma ,%
\mathrm{mag}}=0.013\unit{cm}^{2}\unit{g}^{-1}$, is also much lower than the
values found for the other SNe Ic-BL. \cite{Corsi12} constrained its
progenitor radius to be $R<\left( 1-5\right) R_{\odot }$, consistent with a
compact Wolf-Rayet star. These peculiarities may indicate that PTF10vgv lies
in the gap between SNe Ic and SNe Ic-BL.

The MCMC code can determine the explosion time accurately if the light-curve
data are of high quality. The most excellent case is SN 1998bw, for which
the explosion time was constrained to be $-0.009_{-0.36}^{+0.32}\unit{days}$
relative to the GRB trigger time \citep{WangYu17}.\footnote{%
Previously the burst time of GRB 980426 was constrained to coincide with
that of SN 1998bw within $\left( +0.7,-2.0\right) \unit{days}$ %
\citep[e.g.,][]{Iwamoto99}.} In Table \ref{tbl:exp-time} we compare the
explosion time determined in this work with those given in the literature.
Also listed in this table are the discovery date and date of non-detection.
The explosion time is computed according to the times $T_{\mathrm{start}}$
given in Table \ref{tbl:para}, after correcting for cosmological time
dilation.

In the calculation of the explosion time, we frequently need the time of $V$%
-band maximum, which we consult \cite{Modjaz14}. Because the explosion time
determined in this work is calculated according to the relevant time given
in the original paper, the uncertainties of the explosion time are the
errors given in the original paper, if available, or the errors of $T_{%
\mathrm{start}}$ given in Table \ref{tbl:para}, whichever is larger.

\begin{table*}[tbph]
\caption{Comparison of explosion times derived in this work and previous
papers.}
\label{tbl:exp-time}
\begin{center}
\begin{tabular}{cccccc}
\hline\hline
SN & This work & Previous estimate & Discovery date & Date of non-detection
& References \\ \hline
\multicolumn{1}{l}{1997ef} & 1997/11/25$_{-0.5}^{+0.4}$ & 1997/11/20 & 
1997/11/25 & 1997/11/16 & H97,M00 \\ 
\multicolumn{1}{l}{2002ap} & 2002/01/27$\pm 0.5$ & 2002/01/25.5$\pm 0.5$ & 
2002/01/29 & 2002/01/25 & M02,T06 \\ 
\multicolumn{1}{l}{2003jd} & 2003/10/15.7$\pm 1$ & $<$2003/10/17 & 2003/10/25
& 2003/10/16 & V08 \\ 
\multicolumn{1}{l}{2007bg} & 2007/04/05$\pm 2$ & $-$ & 2007/04/16.15 & 
2007/04/06 & Y10 \\ 
\multicolumn{1}{l}{2007ru} & 2007/11/26$_{-0.6}^{+0.9}$ & 2007/11/25.5 & 
2007/11/27.9 & 2007/11/22 & S09 \\ 
\multicolumn{1}{l}{2009bb} & 2009/03/18$\pm 0.6$ & 2009/03/19.1$\pm 0.6$ & 
2009/03/21.11 & 2009/03/19.2 & P11 \\ 
\multicolumn{1}{l}{2010ah} & 2010/02/20$\pm 1$ & 2010/02/17.8$-$2010/02/23.5
& 2010/02/23.5 & 2010/02/19.4 & C11 \\ 
\multicolumn{1}{l}{2010ay} & 2010/02/23.06$\pm 1.3$ & 2010/02/21.3$\pm 1.3$
& 2010/03/05.45 & 2010/02/17.45 & S12 \\ 
\multicolumn{1}{l}{2012ap} & 2012/02/04.25$\pm 2$ & 2012/02/05$\pm 2$ & 
2012/02/10.23 & $-$ & M15 \\ 
\multicolumn{1}{l}{PTF10qts} & 2010/08/04.6$_{-2}^{+3}$ & $-$ & 2010/08/05.23
& 2010/08/02 & W14 \\ 
\multicolumn{1}{l}{PTF10vgv} & 2010/9/13.2$\pm 0.04$ & $-$ & 2010/09/14.1 & 
2010/09/12.5 & C12 \\ \hline
\end{tabular}%
\end{center}
\par
References: H97: \cite{Hu97}; M00: \cite{Mazzali00}; M02: \cite{Mazzali02};
T06: \cite{Tomita06}; V08: \cite{Valenti08}; S09: \cite{Sahu09}; Y10: \cite%
{Young10}; C11: \cite{Corsi11}; P11: \cite{Pignata11}; C12: \cite{Corsi12};
S12: \cite{Sanders12}; W14: \cite{Walker14}; M15: \cite{Milisavljevic15}; 
\newline
\textbf{Notes.} \newline
UT dates are used in this table. \newline
A hyphen indicates that the date was not specified in the original paper.
\end{table*}

It can be seen from Table \ref{tbl:exp-time} that the times determined in
this work are generally in good agreement with those given in the
literature. The only exception is SN 1997ef, for which our determination,
which is almost coincident with the discovery date, is $\sim 5\unit{days}$
later than that given by \cite{Mazzali00}. Please note that the first
bolometric data point in Figure \ref{fig:1997ef} is $4\unit{days}$ later
than the discovery date.

\subsection{Estimate the appearance of nebular features from early light
curve modeling}

\label{sec:T_Neb}

In this paper we propose to calculate the time $T_{\mathrm{Neb}}$ when the
nebular features begin to emerge in the SN spectra. We searched the spectra
from the references listed in Table \ref{tbl:sample}. The lower limit of $T_{%
\mathrm{Neb}}$ for an SN in question is the latest time at which a spectrum
is photospheric, while the upper limit is the earliest time at which the
spectrum is nebular. We list these constraints in Table \ref{tbl:para}.

The spectrum taken on 26 January, 1998 of SN 1997ef is photospheric %
\citep{Mazzali00}, while the first nebular spectrum is on $+104\unit{days}$
(in rest frame) post \textit{R}-band maximum \citep{Young10}, implying the
transition from photospheric to nebular occurred between these two dates.
Next, the spectrum obtained $+51\unit{days}$ post \textit{B}-band maximum of
SN 2003jd is photospheric, while the spectrum on $+70\unit{days}$ is nebular %
\citep{Valenti08}. For SN~2007bg, the spectrum taken at $+25\unit{days}$
post \textit{R}-band maximum for SN 2007bg is photospheric, while the
spectrum on $+58\unit{days}$ is nebular \citep{Young10}. For SN~2007ru, the
spectrum obtained $70\unit{days}$ after explosion for SN 2007ru is
photospheric, while the spectrum on $200\unit{days}$ is nebular %
\citep{Sahu09}. Next, the spectrum of SN 2009bb on $+45\unit{days}$ past 
\textit{B}-band maximum is photospheric, while the spectrum on $+285\unit{%
days}$ is nebular \citep{Pignata11}. For SN~2012ap, the spectrum taken at $%
+26\unit{days}$ past \textit{B}-band maximum is photospheric, while the
spectrum on $+218\unit{days}$ is nebular \citep{Milisavljevic15}. For
PTF10qts, the spectrum taken $+21\unit{days}$ past \textit{R}-band maximum
is photospheric, while the spectrum on $+230\unit{days}$ is nebular %
\citep{Walker14}. Finally, for PTF10vgv the spectrum obtained at $+35\unit{%
days}$ past \textit{R}-band maximum is photospheric, while the spectrum on $%
+72\unit{days}$ is nebular \citep{Corsi12}.

For SN~2010ah, the spectrum of SN 2010ah on 7, March, 2010 is photospheric %
\citep{Mazzali13}, with no data later than this date being published. The
same situation applies for SN~2010ay, of which that latest spectrum was
obtained on $+24\unit{days}$ past \textit{R}-band maximum is photospheric %
\citep{Sanders12}. For both of these events, the precise timing of the
transition from the photospheric phase to the nebular can only be
constrained to have occurred after these dates.

We can see from Table \ref{tbl:para} that our fitting constraints of $T_{%
\mathrm{Neb}}$ for SNe 2002ap, 2007ru, 2010ah, 2010ay, 2012ap, and PTF10qts
are consistent with observations. For PTF10vgv, $T_{\mathrm{Neb}}$ $\left( 43%
\unit{days}\right) $ is slightly earlier than the lower limit $46.2\unit{days%
}$. For SN 2009bb the given $T_{\mathrm{Neb}}$ $\left( 48\unit{days}\right) $
is $7\unit{days}$ earlier than observation. This may be caused by the helium
envelope of this SN because early-time optical spectra showed evidence for
the presence of helium in this SN (\citealt{Pignata11}; another SN that
evidenced with some helium is SN 2012ap, \citealt{Milisavljevic15}). The
helium envelope will delay the appearance of nebular lines. For SN 2007bg, $%
T_{\mathrm{Neb}}$ $\left( 77\unit{days}\right) $ is $9\unit{days}$ later
than the upper limit. This discrepancy for SN 2007bg might be caused by the
sparsity of data before peak time, see Figure \ref{fig:2003jd}(b).

For 1997ef, $T_{\mathrm{Neb}}$ $\left( 153\unit{days}\right) $ is
significantly later than the appearance of the first nebular spectrum, i.e. $%
119.8\unit{days}$. The situation is less significant but also notable for SN
2003jd, for which we have $92\unit{days}$ versus $81.2\unit{days}$. As
mentioned in Section \ref{sec:gen-imp}, a plausible reason for this large
discrepancy may lie in the failure of our model to fit the early velocity
data of these two SNe. We suggest that a fast-moving shell should be
introduced for SN 1997ef. Such a shell will contribute a significant
fraction of nebular emission and therefore made the appearance of nebular
phase earlier.

In summary, we conclude that $T_{\mathrm{Neb}}$ determined in this way is in
general a good guide for the emergence of nebular features, although it is
not completely accurate. Other factors come into play in determining the
emergence of nebular lines aside from the amount of nebular emission.

\subsection{Correlations}

\label{sec:cor}

In the magnetar model, the required $^{56}$Ni mass is, unsurprisingly,
reduced significantly.\footnote{%
Figure \ref{fig:compare-Lyman} shows that for $t\gtrsim 80\unit{days}$ the
luminosity data calculated according to \cite{Lyman14} are slightly
different from that obtained by integrating individual bands. This will
affect the derived $^{56}$Ni mass. As a result, we call for further study on
the luminosity data for $t\gtrsim 80\unit{days}$.} It is therefore expected
that the $^{56}$Ni mass-energy relation will be quite different, as depicted
in Figure \ref{fig:Ni-E}, where we also plot the ordinary type IIb SN 1993J
and type Ic SN 1994I. It is clear from this figure that the synthesized $%
^{56}$Ni is consistent with ordinary striped envelope SNe. There is no clear
increase of $^{56}$Ni mass with increased energy, contrary to earlier
findings \citep{Mazzali13, Lyman16, Toy16}. In the magnetar+$^{56}$Ni model,
the explosion energies are generally significantly lower than in pure-$^{56}$%
Ni models, regardless whether it is the 1D $^{56}$Ni model or two-component $%
^{56}$Ni model. The explosion energy is no longer the sole decisive factor
for $^{56}$Ni synthesis. In this case the synthesis of $^{56}$Ni may be
determined by other factors, e.g. the radius, and/or density profile of the
progenitor star \citep{Smartt09}. Such diversity may reflect the mass,
binarity, metallicity, mass-loss rate, rotation, and magnetic field of the
main sequence star \citep{Smartt09}.

For an SN Ic-BL, the bolometric corrections at late times can be $\sim 1%
\unit{mag}$\ larger than that at peak times. This indicates that for SNe
1997ef and PTF10vgv, the light curves at late times should be brighter and
flatter than shown in Figures \ref{fig:1997ef}(a) and \ref{fig:2003jd}(f).
As a result, the value of $\kappa _{\gamma ,\mathrm{Ni}}$\ should be larger.
Another impacted parameter is $M_{\mathrm{Ni}}$, although to a less extent.
We estimate that $M_{\mathrm{Ni}}$ may be at most a factor of 2 larger. In
Figure \ref{fig:Ni-E} we show the $^{56}$Ni masses of SNe 1997ef and
PTF10vgv as points connected by horizontal magenta lines with rightward
arrows to indicate the uncertainties introduced by this approximation. As is
clear from Figure \ref{fig:Ni-E}, this does not change our conclusion about
the correlation between $M_{\mathrm{Ni}}$\ and explosion energy. The $^{56}$%
Ni mass of SN 2010ay cannot be constrained, so the uncertainties introduced
by above approximation is irrelevant for SN 2010ay.

Table \ref{tbl:para} shows that the explosion energy of SN 2007ru is quite
low, but the $^{56}$Ni mass $M_{\mathrm{Ni}}$ is not zero. This indicates
that the $^{56}$Ni of this SN was synthesized by the shock wave generated by
the spinning-down magnetar \citep{Nishimura15, Nishimura17, Suwa15}. This is
in agreement with expectations because the rotational energy of the magnetar
powering SN 2007ru is the largest in the SNe Ic-BL sample. The magnetic
field $B_{p}$ is also strong enough to synthesize the needed $^{56}$Ni. In
the magnetar model, both the explosion shock and the magnetar-powered shock
can synthesize $^{56}$Ni. This complicates the $^{56}$Ni mass-energy
relation of SNe Ic-BL.

From Figure \ref{fig:Mej-E} it is clear that the ejecta mass increases with
energy. This is similar to earlier findings \citep{Mazzali13, Lyman16}, but
the energies are much lower than the values given by pure-$^{56}$Ni models %
\citep{Mazzali13, Lyman16}.

We also examined the relation between explosion energy and neutron star
rotational energy, given in Figure \ref{fig:Eexp-Ens}. This figure implies
that there is no clear correlation between these two energies. Figure \ref%
{fig:Bp-E} shows magnetic field $B_{p}$ and energy. No clear correlation is
seen between these two quantities. Because $B_{p}$ harbours a fraction of
the toroidal magnetic field within the neutron star, $B_{p}$ can serve as an
indication of the magnetic energy present within the neutron star. If this
is true, Figure \ref{fig:Bp-E} may imply that the amplification of magnetic
field in the neutron star is unrelated with the explosion energy.

\subsection{Alternative models?}

\label{sec:alter}

In this paper we tested the hypothesis that all SNe Ic-BL are powered by a
combination of input from a magnetar central engine and $^{56}$Ni
synthesized during the initial explosion. Next, we ask the question that can
a pure-$^{56}$Ni model or a pure-magnetar model give comparable results?

\begin{figure*}[tbph]
\centering\includegraphics[width=0.9\textwidth,angle=0]{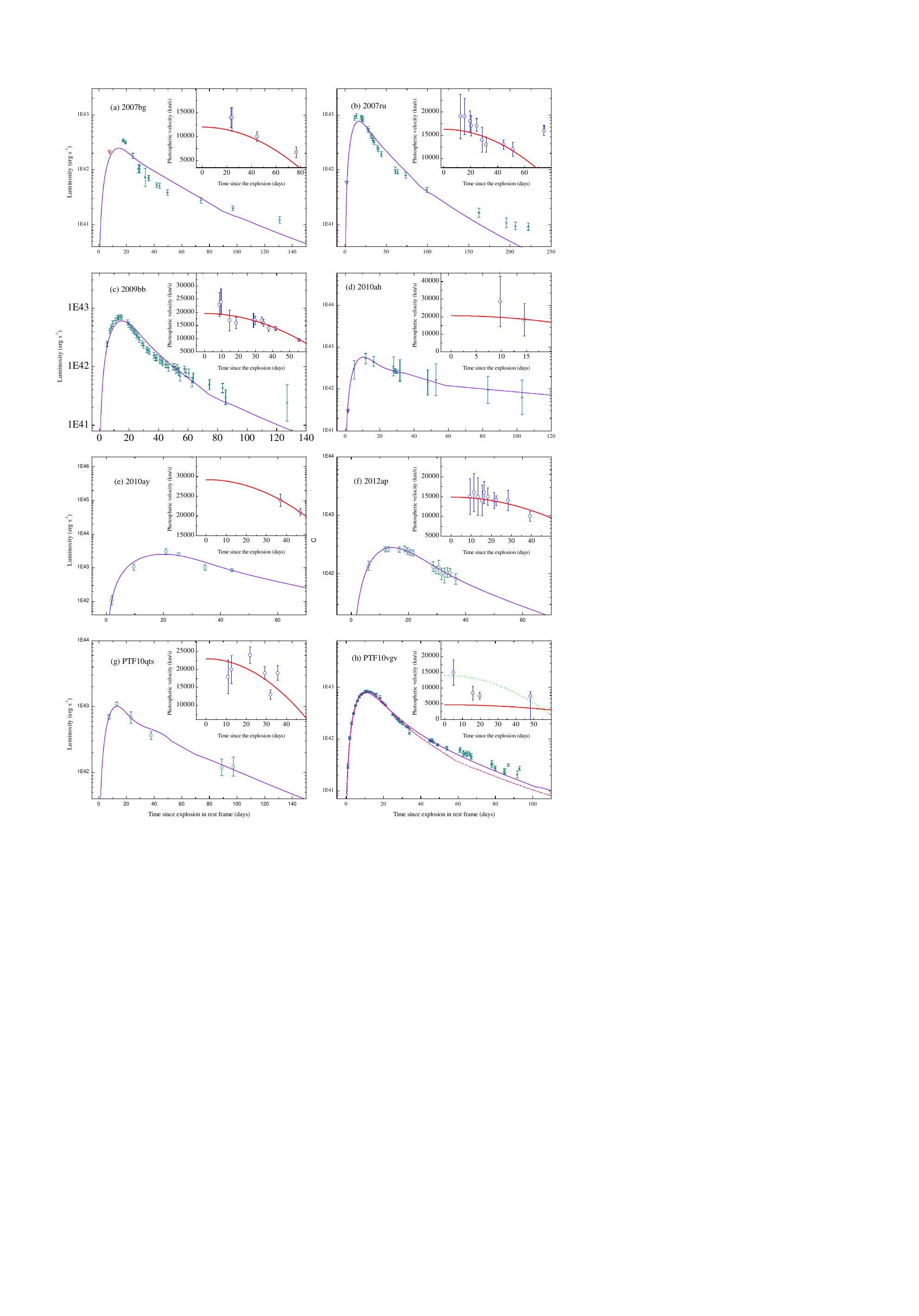}
\caption{Light curves and velocity evolution of the selected SNe reproduced
by the 1D pure-$^{56}$Ni model. The dot-dashed lines in panel (h) are a fit
to the first 50 days of luminosity data by fixing $\protect\kappa _{\protect%
\gamma ,\mathrm{Ni}}=0.027\unit{cm}^{2}\unit{g}^{-1}$.}
\label{fig:56Ni}
\end{figure*}

In Figure \ref{fig:56Ni} we show the best-fit 1D $^{56}$Ni modeling result
for the selected SNe, whose fitting parameters are listed in Table \ref%
{tbl:pure-56Ni}. It is well known that the 1D $^{56}$Ni model cannot give a
satisfactory description for SNe with long observation durations %
\citep{Iwamoto00,Nakamura01a,Maeda03}. As a result the two-component model
was employed to investigate most of the SNe in Figures \ref{fig:1997ef} and %
\ref{fig:2003jd} (SNe 1997ef, 2002ap, \citealt{Maeda03}; SN 2003jd, %
\citealt{Valenti08}; SN 2007bg, \citealt{Young10}). In Figure \ref{fig:56Ni}
we do not show these well-studied SNe (1997ef, 2002ap, 2003jd).

From Figure \ref{fig:56Ni} it can be seen that the 1D $^{56}$Ni model can
only account for the first $\sim 50\unit{days}$\ data (SNe 2010ay, 2012ap).%
\footnote{%
The 1D $^{56}$Ni model can also account for the light curves of SNe 2010ah
and PTF10qts with $\sim 100\unit{days}$ data. However, this may be the
result of the poor quality of the data.} For all well-observed SNe Ic-BL
with observational time $\gtrsim 100\unit{days}$\ (SNe 2007bg, 2007ru,
2009bb in this figure), two-component model should be invoked.

\begin{table*}[tbph]
\caption{Best-fitting parameters for the selected SNe using the 1D $^{56}$Ni
model.}
\label{tbl:pure-56Ni}
\begin{center}
\begin{tabular}{cccccccc}
\hline\hline
SN & $M_{\mathrm{ej}}$ & $M_{\mathrm{Ni}}$ & $v_{\mathrm{sc}0}$ & $T_{%
\mathrm{start}}$ & $\kappa _{\gamma ,\mathrm{Ni}}$ & \multicolumn{2}{c}{$%
\chi ^{2}/\mu $} \\ \cline{7-8}
\multicolumn{1}{l}{} & $\left( M_{\odot }\right) $ & $\left( M_{\odot
}\right) $ & $\left( \unit{km}\unit{s}^{-1}\right) $ & $\left( \unit{days}%
\right) $ & $\left( \unit{cm}^{2}\unit{g}^{-1}\right) $ & pure-$^{56}$Ni
model & magnetar+$^{56}$Ni model \\ \hline
\multicolumn{1}{l}{2007bg} & $1.3\pm 0.5$ & $0.10_{-0.02}^{+0.01}$ & $%
12000_{-1200}^{+1400}$ & $-17.5_{-2.8}^{+4.9}$ & $0.05_{-0.02}^{+0.05}$ & 6.0
& 0.22 \\ 
\multicolumn{1}{l}{2007ru} & $2.6\pm 0.2$ & $0.34\pm 0.01$ & $16300\pm 700$
& $-4.9_{-0.07}^{+0.16}$ & $0.04_{-0.004}^{+0.005}$ & 4.7 & 0.7 \\ 
\multicolumn{1}{l}{2009bb} & $2.5\pm 0.1$ & $0.24\pm 0.007$ & $19600\pm 700$
& $-15.8_{-0.5}^{+0.4}$ & $0.03_{-0.002}^{+0.003}$ & 0.67 & 0.15 \\ 
\multicolumn{1}{l}{2010ah} & $1.0_{-0.5}^{+0.7}$ & $0.17\pm 0.02$ & $%
20600\pm 7000$ & $-2.5_{-0.6}^{+0.9}$ & $\gtrsim 0.5$ & 0.04 & 0.07 \\ 
\multicolumn{1}{l}{2010ay} & $6.4_{-0.5}^{+0.6}$ & $1.2_{-0.07}^{+0.08}$ & $%
27700_{-1600}^{+1800}$ & $1.7_{-0.4}^{+0.3}$ & $-$ & 1.0 & 1.73 \\ 
\multicolumn{1}{l}{2012ap} & $1.6\pm 0.2$ & $0.11\pm 0.004$ & $14900\pm 800$
& $-1.4_{-0.9}^{+0.8}$ & $-$ & 0.12 & 0.14 \\ 
\multicolumn{1}{l}{PTF10qts} & $1.9_{-0.3}^{+0.4}$ & $0.34_{-0.02}^{+0.03}$
& $23000\pm 2000$ & $-12.1_{-1.1}^{+0.9}$ & $0.35_{-0.1}^{+1.2}$ & 1.4 & 2.0
\\ 
\multicolumn{1}{l}{PTF10vgv} & $0.24_{-0.02}^{+0.03}$ & $0.25\pm 0.002$ & $%
4700_{-400}^{+500}$ & $7.1\pm 0.04$ & $0.014\pm 0.001$ & 4.2 & 0.7 \\ 
PTF10vgv-27 & $0.82\pm 0.04$ & $0.26\pm 0.002$ & $14100\pm 500$ & $6.9\pm
0.03$ & $0.027$ & 1.2 & - \\ \hline
\end{tabular}%
\end{center}
\par
\textbf{Notes.} In these fits, we fixed $\kappa =0.1\unit{cm}^{2}\unit{g}%
^{-1}$. For those SNe where $\kappa _{\gamma ,\mathrm{Ni}}$ cannot be
constrained (marked as hyphen), we set $\kappa _{\gamma ,\mathrm{Ni}}=0.027%
\unit{cm}^{2}\unit{g}^{-1}$. For those SNe whose $M_{\mathrm{Ni}}$ cannot be
constrained, i.e. 2010ay and 2012ap, $\chi ^{2}/\mu $ is the result of
pure-magnetar model. PTF10vgv-27 is the best-fitting parameters for the
first 50 days of luminosity data after fixing $\kappa _{\gamma ,{\mathrm{Ni}}%
}=0.027\unit{cm}^{2}\unit{g}^{-1}$.
\end{table*}

We will not examine these SNe within the framework of two-component model in
detail. SN 2010ay in Figure \ref{fig:56Ni} is particularly interesting
because its peak luminosity $\sim 3.0\times 10^{43}\unit{erg}\unit{s}^{-1}$
is comparable to some of the SLSNe, PTF10hgi 
\citep[$3.52\times
10^{43}\unit{erg}\unit{s}^{-1}$;][]{Inserra13}, PTF11rks 
\citep[$4.7\times
10^{43}\unit{erg}\unit{s}^{-1}$;][]{Inserra13}, and PS1-14bj 
\citep[$4.6\times
10^{43}\unit{erg}\unit{s}^{-1}$;][]{Lunnan16}, see also Table 1 in \cite%
{Liu17}. Such SLSNe are usually assumed to be powered by magnetars because
of the failure of $^{56}$Ni model.

The parameters for SN 2010ay are in tension with a typical CCSN. The ratio
of $^{56}$Ni mass to the ejecta mass is 0.19, close to the upper limits 0.2
expected for a CCSN \citep{Umeda08}. The needed $^{56}$Ni mass $1.2M_{\odot
} $\footnote{\cite{Sanders12} estimated the $^{56}$Ni mass of SN 2010ay to
be $0.9\pm 0.1M_{\odot }$. However, if the most luminous data point in the
light curve of SN 2010ay is adopted, a $^{56}$Ni mass of $1.2M_{\odot }$ was
derived \citep{Sanders12}, consistent with our result. \cite{WangWang15b}
found that $M_{\mathrm{Ni}}=2M_{\odot }$ is required to meet the peak
luminosity of SN 2010ay for a $^{56}$Ni model. This higher value results
from adopting a different bolometric correction by \cite{WangWang15b}.} is
higher than that of all SNe Ib/c but SN Ic-BL 2007D 
\citep{Drout11,
Sanders12}. We therefore conclude that SN 2010ay is unlikely to be explained
by a pure-$^{56}$Ni model, including the two-component model.

Another SN that is hard to explain by the $^{56}$Ni model (including the
two-component model) is PTF10vgv because its ejecta mass $M_{\mathrm{ej}%
}=0.24M_{\odot }$\ is smaller than $M_{\mathrm{Ni}}=0.25M_{\odot }$\ (see
Table \ref{tbl:pure-56Ni}). Even if we double its expansion velocity to $%
\sim 10000\unit{km}\unit{s}^{-1}$\ (see Figure \ref{fig:56Ni}h) so that $M_{%
\mathrm{ej}}$\ is doubled, the ratio $M_{\mathrm{Ni}}/M_{\mathrm{ej}}$\ is
still larger than 0.5. The situation becomes even worse for the $^{56}$Ni
model if we take into account the uncertainties in the bolometric light
curve of PTF10vgv. As discussed in Section \ref{sec:cor}, the light curve of
PTF10vgv at late times should be brighter than depicted in Figure \ref%
{fig:56Ni}(h). This indicates that more $^{56}$Ni is needed. The solid lines
in Figure \ref{fig:56Ni}(h) are the best-fitting result allowing $\kappa
_{\gamma ,\mathrm{Ni}}$\ to vary. Table \ref{tbl:pure-56Ni} shows that $%
\kappa _{\gamma ,\mathrm{Ni}}=0.014\unit{cm}^{2}\unit{g}^{-1}$, lower than
the fiducial value $\kappa _{\gamma ,\mathrm{Ni}}=0.027\unit{cm}^{2}\unit{g}%
^{-1}$.\footnote{%
As discussed in Section \ref{sec:cor}, the late-time light curve of SN
1997ef should be slightly flatter than shown in Figure \ref{fig:56Ni}(h).
After taking into account this fact the derived $\kappa _{\gamma ,\mathrm{Ni}%
}$ would be close to the fiducial value.} The dot-dashed lines in Figure \ref%
{fig:56Ni}(h) are the best-fitting result to the first $50\unit{days}$\ of
luminosity data after fixing $\kappa _{\gamma ,\mathrm{Ni}}=0.027\unit{cm}%
^{2}\unit{g}^{-1}$. In this case the initial expansion velocity is $\sim
14000\unit{km}\unit{s}^{-1}$, much higher than the average velocity $\sim
7500\unit{km}\unit{s}^{-1}$\ of this SN. Even with such a high expansion
velocity, the ratio $M_{\mathrm{Ni}}/M_{\mathrm{ej}}=0.32$\ is still higher
than the theoretical upper limit $0.2$. We therefore conclude that PTF10vgv
cannot be explained by $^{56}$Ni model. Recently, it is found iPTF16asu %
\citep{Whitesides17} cannot also be explained by $^{56}$Ni model.

For PTF10qts, \cite{Walker14} obtained $M_{\mathrm{Ni}}=0.35\pm 0.1M_{\odot
} $, based on a model fit to the nebular spectrum of this SN. Such an
estimate of $^{56}$Ni mass is consistent with the value given in Table \ref%
{tbl:pure-56Ni} in the pure-$^{56}$Ni model. This seems to argue against our
hypothesis that all SNe Ic-BL were powered by magnetars. However, on the one
hand, as commented by \cite{Walker14}, any firm conclusions should not be
drawn based on this result because of the low signal-to-noise ratio of the
observed spectrum. On the other hand, for any SN Ic-BL with a long
observational duration, some amount of $^{56}$Ni is indeed required,
although its amount is significantly lower than in the pure-$^{56}$Ni model.

\begin{figure*}[tbph]
\centering\includegraphics[width=0.9%
\textwidth,angle=0]{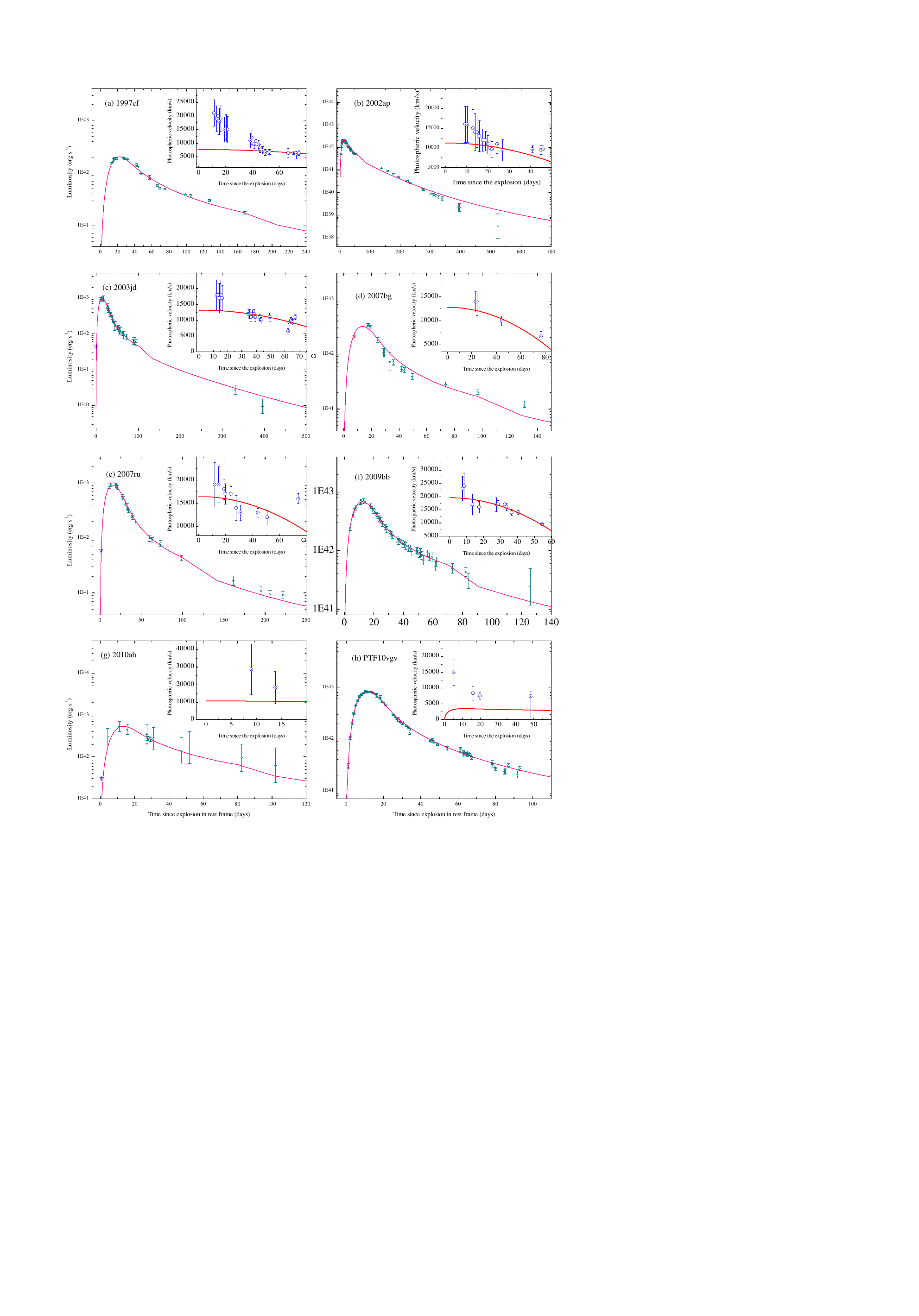}
\caption{Light curves and velocity evolution of the SNe (except for
PTF10qts) in Figures \protect\ref{fig:1997ef} and \protect\ref{fig:2003jd}
reproduced by the pure-magnetar model.}
\label{fig:pure-mag}
\end{figure*}

Now we turn to the pure-magnetar model. For those SNe whose $M_{\mathrm{Ni}}$
can be constrained, the small errors associated with $M_{\mathrm{Ni}}$, as
presented in Table \ref{tbl:para}, clearly indicate the necessity of
including $^{56}$Ni to give the best-fitting results. The synthesis of $%
^{56} $Ni is also expected in a CCSN. Neglecting $M_{\mathrm{Ni}}$ will
usually result in rather poor fitting quality, as shown in Figure \ref%
{fig:pure-mag} with the best-fitting parameters listed in Table \ref%
{tbl:pure-mag}. In Figure \ref{fig:pure-mag} we do not show the fitting
result of SN PTF10qts because the poor data quality always allows for a
\textquotedblleft good" fitting. In Table \ref{tbl:pure-mag} we also compare
the reduced $\chi ^{2}$ of the pure-magnetar model and magnetar+$^{56}$Ni
model.

\begin{table*}[tbph]
\caption{Best-fitting parameters of the SNe in Figures \protect\ref%
{fig:1997ef} and \protect\ref{fig:2003jd} using the pure-magnetar model.}
\label{tbl:pure-mag}
\begin{center}
\begin{tabular}{ccccccccc}
\hline\hline
SN & $M_{\mathrm{ej}}$ & $B_{p}$ & $P_{0}$ & $v_{\mathrm{sc}0}$ & $\kappa
_{\gamma ,\mathrm{mag}}$ & $T_{\mathrm{start}}$ & \multicolumn{2}{c}{$\chi
^{2}/\mu $} \\ \cline{8-9}
\multicolumn{1}{l}{} & $\left( M_{\odot }\right) $ & $\left( 10^{15}\unit{G}%
\right) $ & $\left( \unit{ms}\right) $ & $\left( \unit{km}\unit{s}%
^{-1}\right) $ & $\left( \unit{cm}^{2}\unit{g}^{-1}\right) $ & $\left( \unit{%
days}\right) $ & pure-magnetar & magnetar+$^{56}$Ni \\ \hline
\multicolumn{1}{l}{1997ef} & \multicolumn{1}{l}{$1.8\pm 0.08$} & $1.5\pm
0.02 $ & $37.4\pm 0.2$ & $7674_{-150}^{+149}$ & $\gtrsim 3$ & $%
-25.5_{-0.3}^{+0.4} $ & $2.25$ & $1.51$ \\ 
\multicolumn{1}{l}{2002ap} & \multicolumn{1}{l}{$0.8\pm 0.1$} & $2.3\pm 0.04$
& $48\pm 0.4$ & $11257_{-639}^{+648}$ & $1.5\pm 0.2$ & $-1.9_{-0.3}^{+0.2}$
& $1.3$ & $0.3$ \\ 
\multicolumn{1}{l}{2003jd} & $2.6\pm 0.2$ & $1.8\pm 0.05$ & $20\pm 0.5$ & $%
13134_{-461}^{+476}$ & $1.9_{-0.9}^{+3.2}$ & $-15.9_{-0.6}^{+0.8}$ & $0.25$
& $0.22$ \\ 
\multicolumn{1}{l}{2007bg} & $1.8\pm 0.3$ & $3.1\pm 0.1$ & $38\pm 1$ & $%
12803_{-992}^{+957}$ & $\gtrsim 5$ & $-17\pm 1$ & $1.96$ & $0.22$ \\ 
\multicolumn{1}{l}{2007ru} & $2.7\pm 0.13$ & $1.9\pm 0.01$ & $22\pm 0.15$ & $%
14037_{-538}^{+543}$ & $\gtrsim 5$ & $-4.7\pm 0.1$ & $3.3$ & $3.7$ \\ 
\multicolumn{1}{l}{2009bb} & $2.2\pm 0.1$ & $2.4\pm 0.06$ & $28\pm 0.4$ & $%
19540_{-757}^{+616}$ & $4.9_{-2.8}^{+3.3}$ & $-12.8\pm 0.3$ & $0.20$ & $0.15$
\\ 
\multicolumn{1}{l}{2010ah} & $0.76_{-0.45}^{+1.40}$ & $1.4_{-0.9}^{+0.6}$ & $%
29_{-8}^{+3}$ & $21879_{-8915}^{+12335}$ & $\gtrsim 2.7$ & $%
-2.4_{-0.7}^{+1.0}$ & $0.07$ & $0.07$ \\ 
PTF10qts & $2.7_{-0.7}^{+2.0}$ & $1.1\pm 0.2$ & $20_{-2}^{+1}$ & $%
20478_{-1910}^{+2421}$ & $\gtrsim 1$ & $-13.5\pm 2$ & $1.7$ & $2.0$ \\ 
\multicolumn{1}{l}{PTF10vgv} & $0.29_{-0.007}^{+0.008}$ & $%
2.4_{-0.014}^{+0.016}$ & $24\pm 0.2$ & $\sim 0$ & $4.2_{-2.7}^{+2.5}$ & $%
6.9_{-0.03}^{+0.04}$ & $1.2$ & $0.7$ \\ \hline
\end{tabular}%
\end{center}
\par
\textbf{Notes.} In these fits, we fixed $\kappa =0.1\unit{cm}^{2}\unit{g}%
^{-1}$.
\end{table*}

The high fitting quality of the magnetar+$^{56}$Ni model can be most easily
appreciated by comparing the reproduced light curves of SN 2002ap by these
two models. The situation of SN 2002ap in the pure-magnetar model is similar
to SN 1998bw in this same model, see Figure 8 of \cite{Moriya17}. The ejecta
masses given by the pure-magnetar model are frequently unreasonable, e.g.
SNe 2002ap, 2010ah and PTF10vgv. For SN 1997ef, the magnetar+$^{56}$Ni model
is favored not only because of the smaller reduced $\chi ^{2}$ compared to
the pure-magnetar model, but also because of the broad peak of this SN, as
found by \cite{Iwamoto00}. Comparing Figures \ref{fig:1997ef}(a) and \ref%
{fig:pure-mag}(a) indicates that the magnetar+$^{56}$Ni model captures the
broad peak of this light curve better than the pure-magnetar model, see
Figure 1 in \cite{WangHan16} for a clearer rendering.\footnote{%
We note that the two-component model cannot capture the broad peak of SN
1997ef, see the inset (circles versus dotted line) of Figure 4 of \cite%
{Maeda03}. This indicates that the magnetar+$^{56}$Ni model is the best to
account for SN 1997ef.}

For SNe 2003jd, 2007ru, 2010ah, and PTF10vgv, the contribution of $^{56}$Ni
is not necessary to give an acceptable fitting result. However, the velocity
fitting result of PTF10vgv (see Figure \ref{fig:pure-mag}) is not good
enough. The low velocities are required for PTF10vgv in the pure-magnetar
model because of the slow decline rate of the light curve.

The above results indicate that the magnetar+$^{56}$Ni model is the best
model in reproducing the light curves and velocity evolutions of the SNe
Ic-BL sample, although some SNe can also be described by the two-component $%
^{56}$Ni model (e.g., SN 2002ap), while some others (e.g., SNe 2003jd,
2007ru, and 2010ah) can also be well reproduced by the pure-magnetar model.

We note that different models usually give different explosion times, as can
be found by comparing $T_{\mathrm{start}}$ presented in Tables \ref{tbl:para}%
, \ref{tbl:pure-56Ni}, and \ref{tbl:pure-mag}. Tables \ref{tbl:para} and \ref%
{tbl:pure-mag} show that the explosion times determined by the magnetar+$%
^{56}$Ni model are all later than that determined by the pure-magnetar
model, with PTF10vgv the only exception. This can be well understood. To
account for the late-time light curves, the spin-down timescales of the
magnetars in the pure-magnetar model have to be longer than in the magnetar+$%
^{56}$Ni model. This will result in slow rise rate and therefore the
explosion times must be somewhat earlier. The slow rise rates in the
pure-magnetar model suffer from some tension with the upper limits of the
light curves of SNe 2003jd, 2007bg, 2007ru, as can be seen from Figure \ref%
{fig:pure-mag}.

Comparison of $T_{\mathrm{start}}$ in Tables \ref{tbl:para} and \ref%
{tbl:pure-56Ni} shows that the explosion times in the magnetar+$^{56}$Ni
model are usually later than that in the $^{56}$Ni model, except for
PTF10vgv. This can be understood by comparing the spin-down timescale of the
magnetar, $\tau _{\mathrm{sd}}$, with the $^{56}$Ni decay timescale, $\tau _{%
\mathrm{Ni}}=8.8\unit{days}$. It is found that the spin-down timescales of
the magnetars powering these SNe are all shorter than $\tau _{\mathrm{Ni}}$,
with only one exception, $\tau _{\mathrm{sd}}\left( \mathrm{PTF10vgv}\right)
=15.9\unit{days}$. If $\tau _{\mathrm{sd}}<\tau _{\mathrm{Ni}}$, the energy
of the magnetar is released more rapidly in the magnetar model than in the $%
^{56}$Ni model, the rise time in the magnetar model is shorter than in the $%
^{56}$Ni model. This is why the explosion time of PTF10vgv in the magnetar
model is earlier than in the $^{56}$Ni model because in this case $\tau _{%
\mathrm{sd}}>\tau _{\mathrm{Ni}}$. By this way, it is not difficult to
understand why the explosion time of SN 1997ef is almost coincident with the
discovery date in the magnetar+$^{56}$Ni model. From Table \ref{tbl:para} it
is found that $\tau _{\mathrm{sd}}=0.01\unit{days}$ for this SN. The energy
was almost explosively released.

\section{Conclusions}

\label{sec:conclusion}

The mechanism for the formation of SNe Ic-BL is still unclear. Recently
there is evidence that SNe Ic-BL are powered by magnetars \citep{WangYu17}.
Indeed, for all of the SNe Ic-BL that were observed to phases $\gtrsim 300%
\unit{days}$ when the contribution from $^{56}$Ni decays significantly,
there is evidence for magnetar formation.

Motivated by this evidence, we studied a sample of $N=11$ SNe Ic-BL and
obtain their light curve fitting parameters. From this study it is evident
that the sample of SNe Ic-BL can be reasonably described by the magnetar+$%
^{56}$Ni model. The magnetar+$^{56}$Ni model naturally reduces the needed $%
^{56}$Ni and simultaneously accounts for the origin of the huge kinetic
energies observed in SNe Ic-BL, with only one exception, SN 2010ay, whose
large explosion energy could be attributed to the large photometric
uncertainties. We also examine the possibility for the pure-$^{56}$Ni or
pure-magnetar model to explain the light curve and velocity evolution. It is
found that SNe 2010ay, PTF10vgv, and iPTF16asu (3 out of 12) are unlikely
explained by the (two-component) $^{56}$Ni model, while some SNe 2003jd,
2007ru, and 2010ah (not all in the sample) are compatible with the
pure-magnetar model.

Our results indicate that the synthesized $^{56}$Ni mass does not increase
with explosion energy or neutron star rotational energy. The $^{56}$Ni mass
is consistent with ordinary SNe. The relation between magnetic field and
explosion energy seems to indicate that the amplification of magnetic field
of the neutron star is independent of the explosion energy. To get a more
robust statistical result, more high-quality observations are definitely
needed.

\begin{acknowledgements}
We thank the anonymous referee for constructive suggestions.
We are grateful to Maryam Modjaz and Yuqian Liu for sending us the photospheric velocity
 data before the publication of their paper.
This work is supported by the National Program on Key Research and Development Project
 of China (Grant No. 2016YFA0400801), National
Basic Research Program of China (\textquotedblleft 973" Program, Grant
No. 2014CB845800) and the National Natural Science Foundation of China (grant
Nos. U1331202, 11533033, U1331101, 11673006, 11573014, 11422325 and 11373022).
D.X. acknowledges the support of the
One-Hundred-Talent Program from the National Astronomical Observatories,
Chinese Academy of Sciences), and by the Strategic Priority Research
Program ``Multi-wavelength Gravitational Wave Universe" of the CAS (No.
XDB23040100).
\end{acknowledgements}

\end{document}